\pdfoutput=1
\documentclass[PAPER, coverpage=false, UKenglish, atlasdraft=false, cernpreprint, texmf, orcidlogo]{atlasdoc}
\usepackage{atlaspackage}
\usepackage{atlasbiblatex}

\usepackage{atlasphysics}
\graphicspath{{logos/}{figures/}}

\usepackage{ANA-HMBS-2024-04-PAPER-defs}
\usepackage{multirow}
\usepackage{makecell}
\usepackage{subfig}
\usepackage{tabu}
\usepackage{atlasbsm}

\newcommand{\AtlasCoordFootnote}{%
ATLAS uses a right-handed coordinate system with its origin at the nominal interaction point (IP)
in the center of the detector and the $z$-axis along the beam pipe.
The $x$-axis points from the IP to the center of the LHC ring,
and the $y$-axis points upwards.
Cylindrical coordinates $(r,\phi)$ are used in the transverse plane,
$\phi$ being the azimuthal angle around the $z$-axis.
The pseudorapidity is defined in terms of the polar angle $\theta$ as $\eta = -\ln \tan(\theta/2)$.
Angular distance is measured in units of $\Delta R \equiv \sqrt{(\Delta\eta)^{2} + (\Delta\phi)^{2}}$.
The distance from the interaction point along the $z$-axis is denoted as $|z|$.
}

\newcommand*{\chinoonemp}{\ensuremath{\mathchoice%
{\displaystyle\raise.4ex\hbox{\(\displaystyle\tilde\chi^\mp_1\)}}%
{\textstyle\raise.4ex\hbox{\(\textstyle\tilde\chi^\mp_1\)}}%
{\scriptstyle\raise.3ex\hbox{\(\scriptstyle\tilde\chi^\mp_1\)}}%
{\scriptscriptstyle\raise.3ex\hbox{\(\scriptscriptstyle\tilde\chi^\mp_1\)}}}\xspace}

\AtlasTitle{Reinterpretation of searches for supersymmetry in models with variable R-parity-violating coupling strength using the full ATLAS Run 2 Dataset}

\AtlasAbstract{%
A collection of thirteen ATLAS searches for supersymmetry (SUSY) models, optimized for R-parity-conserving (RPC) and R-parity-violating (RPV) SUSY, are reinterpreted in SUSY models with variable RPV coupling strength, which determines whether the lightest supersymmetric particle decays promptly or is long-lived. The dataset corresponds to an integrated luminosity of 140~fb$^{-1}$ of proton–proton collisions at a centre-of-mass energy of $\sqrt{s}=13$ TeV collected between 2015 and 2018 by the ATLAS detector at the Large Hadron Collider. Limits are set at 95\% confidence level on the mass of pair-produced gluinos decaying to final states enhanced or depleted with top quarks, and on the masses of pair-produced top squarks, tau-sleptons, or charginos and neutralinos. In a model of pair-produced gluinos decaying to final states enhanced with top quarks, a lower limit of 1.8~TeV on the gluino mass is set regardless of the RPV coupling value. In the gluino model with decays to first and second generation quarks, gluino masses are excluded up to 1.6--2.2 (1.6--2.5)~TeV for different values of the RPV coupling $\lambda^{'}$ ($\lambda^{''}$). Top-squark masses up to 2.4~TeV are excluded at high values of $\lambda^{''}$, compared to 1.0--1.7 TeV for low and intermediate $\lambda^{''}$. Tau-slepton masses between 180~GeV and 340~GeV are excluded for $\lambda$ couplings smaller than $10^{-4}$. Higgsino masses up to 800 GeV--1.0 TeV are excluded when $\lambda_{i33}$ is larger than $4\times10^{-5}$. This work extends the analyses beyond RPC scenarios to a broad class of RPV frameworks and achieves significantly improved sensitivity to a diverse range of long-lived particle signatures.
}

\AtlasRefCode{HMBS-2024-04}

\PreprintIdNumber{CERN-EP-2026-054}

\AtlasJournal{Journal of High Energy Physics}
\AtlasCoverEgroupAnalysisTeam{atlas-hmbs-2024-04-analysis-team@cern.ch}

\hypersetup{pdftitle={ATLAS document},pdfauthor={The ATLAS Collaboration}}

\begin{document}

\maketitle

\section{Introduction}


Supersymmetry (SUSY)~\cite{Miyazawa:1966,Ramond:1971gb,Golfand:1971iw,Neveu:1971rx,Neveu:1971iv,Gervais:1971ji,Volkov:1973ix,Wess:1973kz,Wess:1974tw}
is a generalization of space-time symmetries that extends the Standard Model (SM) by introducing supersymmetric partners for every SM particle with
identical quantum numbers except for a half-unit difference in spin.
The scalar partners of the left- and right-handed quarks, the squarks $\tilde{q}_L$ and $\tilde{q}_R$, mix to form two mass eigenstates $\tilde{q}_1$ and $\tilde{q}_2$.
Similarly, the scalar partners of the left- and right-handed charged (neutral) leptons are known as sleptons (sneutrinos) and mix to form two mass eigenstates $\tilde{l}_1$ and $\tilde{l}_2$.
Superpartners of the charged and neutral electroweak and Higgs bosons, so-called winos, binos and higgsinos, also
mix producing charginos ($\tilde{\chi}^{\pm}_{1,2}$) and neutralinos ($\tilde{\chi}^{0}_{1,2,3,4}$) with subscripts indicating increasing mass. In this paper, the lightest neutralino, \ninoone,
is assumed to be the lightest supersymmetric particle (LSP).
Squarks and the fermionic partners of the gluons, the gluinos ($\tilde{g}$), could be produced in strong-interaction processes at the Large Hadron Collider~\cite{Evans:2008zzb} (LHC) with large cross-sections.

In the most generic superpotential~\cite{Martin:1997ns}, the following Yukawa and bilinear couplings can lead to baryon- and
lepton-number violation:
\begin{equation}
\mathcal{W}_{\text{RPV}} = \frac{\lambda_{ijk}}{2} L_i L_j \bar{E}_k + \lambda'_{ijk} L_i Q_j \bar{D}_k +  \frac{\lambda''_{ijk}}{2} \bar{U}_i \bar{D}_j \bar{D}_k + \kappa_i L_i H_u ,
\label{eq:RPV_lagrangian}
\end{equation}
where $i$, $j$, and $k$ are quark and lepton generation indices. The third term corresponds to a baryon-number-violating term, while the others correspond to lepton-number-violating terms. The $L_i$ and $Q_i$ represent the lepton and quark
SU(2)$_\text{L}$ doublet superfields and $H_u$ the Higgs superfield that couples to up-type quarks. The $\bar{E}_i$, $\bar{D}_i$ and $\bar{U}_i$
are the lepton, down-type quark and up-type quark SU(2)$_\text{L}$ singlet superfields, respectively.
The couplings are $\lambda$, $\lambda'$, $\lambda''$, and $\kappa$.
The $\lambda$ and $\lambda''$ couplings are antisymmetric under the exchange of $i \leftrightarrow j$ and $j \leftrightarrow k$, respectively.
While the terms in Eq.~(\ref{eq:RPV_lagrangian}) are removed in many SUSY scenarios by imposing an additional $\mathbb{Z}_2$ symmetry ($R$-parity)~\cite{Fayet:1976et},
there is no fundamental theoretical reason to do so.
Conversely, $R$-parity-violating (RPV) scenarios, in which at least some of these RPV couplings are non-zero, are also well motivated~\cite{Dreiner:1997uz,Barbier:2004ez}.
While searches for $R$-parity-conserving (RPC) models have set stringent constraints on the masses of SUSY particles, the same does not hold for RPV scenarios~\cite{Evans:2013jna,Batell:2013zwa}.
Non-zero RPV couplings violate baryon and/or lepton number conservation and lead to distinctive phenomenological consequences~\cite{Barbier:2004ez}.
Some of these couplings may even offer potential solutions to commonly discussed shortcomings of the Standard Model.
The $\lambda$ and $\lambda'$ couplings can generate neutrino masses and mixing~\cite{Hall:1983id,Babu:1989px}, while $\lambda'$ and $\lambda''$ affect low-energy observables~\cite{Allanach:1999ic,Domingo:2018qfg}.
The $\lambda''$ terms can also provide mechanisms for baryogenesis~\cite{Dimopoulos:1987rk,Adhikari:1996mc} and arise naturally in minimal flavor violation SUSY~\cite{Csaki:2011ge}.
RPV scenarios may impact the value of muon $(g-2)_\mu$~\cite{Chakraborty:2015bsk,Altmannshofer:2020axr} and cause deviations in flavor-related observables~\cite{Trifinopoulos:2018rna,Altmannshofer:2020axr}.

If $R$-parity is conserved, SUSY particles are produced in pairs and decay either directly or via cascades to the LSP that is stable and escapes the detector unseen.
Introducing non-zero RPV couplings renders the LSP unstable and opens decay channels to fermion-sfermion pairs with the subsequent decay of the sfermion into SM particles through the relevant RPV coupling shown in Eq.~(\ref{eq:RPV_lagrangian})~\cite{Diglio:2017stops}.
The LSP lifetime, \tauLSP, depends on the RPV coupling strength as well as the masses of the sfermions involved in the decay.
In the parameter space of small RPV couplings (\textasciitilde $10^{-2}$), the LSP can become long-lived and decay after traversing
a measurable distance within the detector volume. In the limit where the RPV coupling is vanishingly small (\textasciitilde $10^{-5}$), the majority of LSP decays occur
outside the detector volume, producing the same experimental signature as RPC SUSY. If the RPV coupling is increased beyond the regime of prompt LSP decays, then the $\tau$-sleptons, squarks, gluinos and $\tilde{\chi}_2^0$'s can decay directly to SM particles via the large RPV coupling. Furthermore, non-zero coupling values allow for single sparticle production modes. Thus, different RPV coupling strengths lead to several distinct regimes of SUSY phenomenology.

Most existing SUSY searches assume either $R$-parity conservation or RPV couplings that are large enough to yield prompt LSP decays, leaving the intermediate regime - where the LSP is long-lived - largely unconstrained.
This paper reinterprets published searches for SUSY in ATLAS to derive mass limits across the full range of RPV coupling strengths. Models with variable baryon- or lepton-number-violating RPV couplings $\lambda$, \lambdap or \lambdapp are considered. The \texttt{RECAST}~\cite{Cranmer:2018cqv} framework, widely used by ATLAS ~\cite{SUSY-2020-05,SUSY-2020-09,SUSY-2020-15,EXOT-2023-01}, is employed for some reinterpretations, while others directly reuse the original analysis code.
In all cases, the full ATLAS detector simulation based on {\scshape Geant4}~\cite{Agostinelli:2002hh} is used, which is crucial when reinterpreting prompt searches with long-lived particle (LLP) signals, as LLP reconstruction efficiencies and event topologies cannot be reliably captured using higher-level or generator-level approximations~\cite{deFavereau:2013fsa,Bierlich:2024vqo}.
Furthermore, since the signal efficiency depends on the displacement of the LLP decay vertex, the associated systematic uncertainties are evaluated as a function of the vertex displacement.
While the signal efficiency and associated systematics are specifically adapted, all other aspects of the analyses - object definitions, background estimates, statistical models, and likelihoods - are taken directly from the original publications.
Earlier reinterpretation of SUSY searches with RPV SUSY models was performed using the Run-1~\cite{ATLAS-CONF-2015-018} data. This work extends the effort to the full Run-2 dataset, covers a broader set of RPV SUSY models, and includes both prompt and long-lived particle searches.
More generally, these reinterpretations demonstrate that searches targeting specific benchmark models retain sensitivity to a broad range of scenarios beyond those for which they were explicitly optimized.

The paper is structured as follows:
Section~\ref{sec:detector} provides a brief description of the ATLAS detector.
Section~\ref{sec:susymodels} describes the signal models considered, including the existing external constraints and the generation of signal samples.
The analyses used to constrain these models are summarized in Section~\ref{sec:analyses}.
Section~\ref{sec:obj_and_uncert} presents the study of systematic uncertainties associated with the application of standard reconstruction algorithms to non-prompt objects.
Finally, the results are presented in Section~\ref{sec:result}.


%

%
\section{ATLAS detector}
\label{sec:detector}

%
%
%
%
%
%
%
%
%
%
%

%

The ATLAS experiment~\cite{PERF-2007-01} at the LHC is a multipurpose particle detector
with a forward--backward symmetric cylindrical geometry and a near \(4\pi\) coverage in
solid angle.\footnote{\AtlasCoordFootnote}
It consists of an inner tracking detector surrounded by a thin superconducting solenoid
providing a \qty{2}{\tesla} axial magnetic field, electromagnetic and hadronic calorimeters, and a muon spectrometer.
The inner tracking detector covers the pseudorapidity range \(|\eta| < 2.5\).
It consists of silicon pixel, silicon microstrip, and transition radiation tracking detectors.
Lead/liquid-argon (LAr) sampling calorimeters provide electromagnetic (EM) energy measurements
with high granularity within the region \(|\eta|< 3.2\).
A steel/scintillator-tile hadronic calorimeter covers the central pseudorapidity range (\(|\eta| < 1.7\)).
The endcap and forward regions are instrumented with LAr calorimeters
for EM and hadronic energy measurements up to \(|\eta| = 4.9\).
The muon spectrometer surrounds the calorimeters and is based on
three large superconducting air-core toroidal magnets with eight coils each.
The field integral of the toroids ranges between \num{2.0} and \qty{6.0}{\tesla\metre}
across most of the detector.
The muon spectrometer includes a system of precision tracking chambers up to \(|\eta| = 2.7\) and fast detectors for triggering up to \(|\eta| = 2.4\).
The luminosity is measured mainly by the LUCID--2~\cite{LUCID2} detector which is located close to the beampipe.
A two-level trigger system was used to select events~\cite{TRIG-2016-01}.
The first-level trigger is implemented in hardware and used a subset of the detector information
to accept events at a rate close to \qty{100}{\kHz}.
This is followed by a software-based trigger that
reduced the accepted rate of complete events to \qty{1.25}{\kHz} on average
depending on the data-taking conditions.
A software suite~\cite{SOFT-2022-02} is used in data simulation, in the reconstruction
and analysis of real and simulated data, in detector operations, and in the trigger and data acquisition
systems of the experiment.


%

%
%

%
\section{SUSY models}
\label{sec:susymodels}

The sensitivity of a suite of ATLAS searches is evaluated on a set of simplified SUSY models~\cite{Alwall:2008ve,Alwall:2008ag,Alves:2011wf}. The details of these models, the existing external constraints, and the signal sample generation are described in the following section.

\subsection{Model parameters}

Six simplified RPV SUSY models are considered in this paper. The models are chosen to span both strong and electroweak production modes and to cover representative decay topologies and lifetimes, with particular emphasis on higgsino- and top-squark-mediated scenarios which are motivated by naturalness~\cite{Papucci:2011wy}. Within each set of \lambdaijk, \lambdapijk, \lambdappijk couplings, typically only one coupling is considered to be non-zero in each simplified model (with the exception of the $\lambda'_{111}$ scenario, where both $\lambdap_{111}$ and $\lambdap_{211}$ are non-zero and identical). The choice of having only one non-zero baryonic or leptonic coupling is made for simplicity.

The LSP is assumed to be either bino-like or higgsino-like~\cite{Martin:1997ns,SUSY-2023-10}.
In the models with $\lambdapp$ couplings, the LSP mass is set to 200 \GeV, which allows for LSP decays into top quarks. For the other models, the LSP masses are chosen to maximize the exclusion reach of the pair-produced sparticle mass in the analysis targeting the corresponding RPC scenarios. %
The choice of a bino-like neutralino for most of the models is made for simplicity, as the absence of a chargino in the particle spectrum reduces the number of possible squark or gluino decays.
Moreover, the nature of the neutralino has an impact on its lifetime, e.g. a higgsino-like neutralino has a shorter lifetime than a bino-like neutralino due to the large Yukawa coupling to the top squark.

\begin{table}[htbp]
\begin{center}
\caption{Summary of signal models. The symbols, $q$, $\ell$, and $\nu$ denote first and second generation quarks ($u$, $d$, $s$, $c$), light charged leptons ($e$, $\mu$), and Standard Model neutrinos, respectively. First and second generation squark masses are assumed to be degenerate
\mbox{($\tilde{q}$ = $\tilde{u}$, $\tilde{d}$, $\tilde{s}$,  $\tilde{c}$)}.
Left- and right-handed superpartner masses are also assumed to be degenerate
($\tilde{q}= \tilde{q}_1, \tilde{q}_2$),
except for the stop model where the right-handed top-quark partner is assumed to be lighter.
Within each cell of the ``Decay'' column, the decay modes are listed from top to bottom in order of increasing values of the RPV coupling.
}
\label{tab:Modelsummary}
\setlength{\tabcolsep}{0.6pc}{\renewcommand{\arraystretch}{1.4}
\begin{tabular}{| c | c | c | c | c |}
\hline
\hline
\multirow{2}{*}{Model name}  		& \multirow{2}{*}{Coupling}  & \multirow{2}{*}{Decay}	& \multirow{2}{*}{\begin{minipage}{2.5cm}Other \newline sparticle masses \end{minipage}}  & \multirow{2}{*}{LSP} \\
&           & 	    &  &  \\
\hline
\multirow{3}{*}{Gqq+LQD}	& \multirow{3}{*}{$\lambdap_{111}$ = $\lambdap_{211}$}	& $\gluino \rightarrow qq \ninoone$  & $m(\tilde{q})=3$ \tev	& \multirow{3}{*}{\begin{minipage}{3.0cm}bino-like \ninoone, \\ $m(\ninoone)=500$ \gev \end{minipage}}  \\
&  & $\gluino \rightarrow qq \ninoone(\rightarrow \ell q q'/\nu q q')$  & $m(\tilde{t},\tilde{b})=5$ \tev &  \\
&  & $\gluino \rightarrow \ell q q'/\nu q q'$  &  &  \\
\hline
\multirow{3}{*}{Gqq+UDD}	& \multirow{3}{*}{$\lambdapp_{112}$}	& $\gluino \rightarrow qq \ninoone$  & $m(\tilde{q})=3$ \tev	& \multirow{9}{*}{\begin{minipage}{3.0cm}bino-like \ninoone, \\ $m(\ninoone)=200$ \gev \end{minipage}}  \\
&  & $\gluino \rightarrow qq \ninoone(\rightarrow qqq)$  & $m(\tilde{t},\tilde{b})=5$ \tev &  \\
&  & $\gluino \rightarrow qqq$  &  &  \\
\cline{1-4}
\multirow{3}{*}{Gtt}	& \multirow{3}{*}{$\lambdapp_{323}$}	& $\gluino \rightarrow tt \ninoone$  & $m(\tilde{q})=5$ \tev	&    \\
&  & $\gluino \rightarrow tt \ninoone(\rightarrow tbs)$  & $m(\tilde{t},\tilde{b})=3$ \tev &  \\
&  & $\gluino \rightarrow tbs$  &  &  \\

\cline{1-4}
\multirow{3}{*}{Stop}	& \multirow{3}{*}{$\lambdapp_{323}$}	& $\stopone \rightarrow t\ninoone$  & $m(\tilde{q},\tilde{g})=3$ \tev	&    \\
&  & $\stopone \rightarrow t\ninoone(\rightarrow tbs)$  & $m(\tilde{t}_2,\tilde{b})=3$ \tev &  \\
&  & $\stopone \rightarrow bs$  &  &  \\
\hline
\multirow{6}{*}{Stau} & \multirow{6}{*}{\shortstack[c]{\(\lambda_{i33}\),\\ \(i \in \{1,2\}\)}} & $\stau \rightarrow \tau \ninoone$ & m($\slepton$) = 5 \TeV & \multirow{6}{*}{\begin{minipage}{3.0cm}bino-like \ninoone, \\ $m(\ninoone)=50$ \gev \end{minipage}}\\ %
& & $\stau \rightarrow \tau \ninoone(\rightarrow \ell \tau \nu/\tau \tau \nu)$ & &  \\
& & $\stau \rightarrow \tau \nu$ & &  \\
\cline{3-3}
& & $\snu_\tau \rightarrow \nu \ninoone$ & &  \\
& & $\snu_\tau \rightarrow \nu \ninoone(\rightarrow \ell \tau \nu/\tau \tau \nu)$ & &  \\
& & $\snu_\tau \rightarrow \tau \ell$ & &  \\

\hline
\multirow{6}{*}{Higgsino}	& \multirow{6}{*}{\shortstack[c]{$\lambda_{i33}$, \\ $i \in \{1,2\}$}}	& $\chinoonepm \rightarrow q q' \ninoone / \ell\nu \ninoone$  & m($\stau$) = 3 \TeV	& \multirow{6}{*}{\begin{minipage}{3.0cm}higgsino-like \ninoone, \\ $\Delta$m(\chinoonepm, \ninoone) = 0.25 \GeV  \\ $\Delta$m(\ninotwo, \ninoone) = 0.5 \GeV \end{minipage}}\\%
&  & $\chinoonepm \rightarrow q q' \ninoone / \ell\nu \ninoone(\rightarrow \ell \tau \nu / \tau \tau \nu)$  & m($\slepton$) = 5 \TeV &  \\
&  & $\chinoonepm \rightarrow \ell\tau\tau/\ell\nu\nu/\tau\nu\nu $  &  &  \\
\cline{3-3}
&  & $\ninotwo \rightarrow q q \ninoone / \ell \ell \ninoone$  &  &  \\
&  & $\ninotwo \rightarrow q q \ninoone / \ell \ell \ninoone(\rightarrow \ell \tau \nu / \tau  \tau \nu)$  &  &  \\
&  & $\ninotwo \rightarrow \ell \tau \nu /\tau \tau \nu $  &  &  \\
\hline
\hline
\end{tabular}
}
\end{center}

\end{table}

Despite the use of simplified models, the masses of all squarks or sleptons have to be specified even if they are not considered in the accessible sparticle spectrum, since the LSP lifetime depends on the choice of these masses. The exclusion limits established in this paper are presented as a function of the RPV coupling strength, the LSP lifetime, and the branching fraction of the pair-produced sparticles. The relation between the coupling strength and the LSP lifetime and branching fraction is determined by the choice of relevant squark or slepton masses: for example, the estimated mean decay length for a bino-like LSP neutralino moving with speed $\beta$ and Lorentz factor $\gamma$, decaying via a \lambdapp coupling is ~\cite{Dreiner:1991pe}

\begin{equation}
L[cm] = \frac{0.9\beta\gamma}{{\lambdapp}^2}\left(\frac{m(\tilde{q})}{100~\gev} \right)^4 \left(\frac{1~\gev}{m(\ninoone)} \right)^5.
\label{eq:lifetime}
\end{equation}

The squark and slepton masses are set to be large enough to evade constraints on direct squark- and slepton-pair production, but small enough to ensure that the LSP becomes promptly decaying once the direct RPV decays of the pair-produced sparticles become non-negligible. %
The computation of lifetime and branching fractions is performed with \textsc{SPheno 4.0.2}~\cite{Porod:2003um,Porod:2011nf} in combination with \textsc{SARAH 4.12.0}~\cite{Staub:2013tta}.

The simplified models considered in this paper are described below, with a summary provided in Table~\ref{tab:Modelsummary}.
\begin{description}

\item[Gqq+LQD model:] This model contains light gluinos and a 500~GeV bino-like $\ninoone$  LSP, together with non-zero $\lambdap_{111}$ and $\lambdap_{211}$ couplings where $\lambdap_{111} = \lambdap_{211}$, which corresponds to interactions involving electrons and muons, while all other RPV couplings are set to zero. The gluinos are pair-produced and decay via off-shell first- and second-generation squarks. As shown in Figure \ref{fig:diagrams1p1}, for low values of the RPV coupling, the gluino decays as $\gluino \rightarrow q q \ninoone$ \mbox{($q$ = $u$, $d$, $s$, $c$)}
with the subsequent neutralino decay proceeding via the LQD operator as $\ninoone \rightarrow \ell q q' /\nu q q'$ ($\ell=e,\mu$)
thus yielding final states containing both jets and leptons. For larger values of the coupling, the gluino can also decay directly as $\gluino \rightarrow \ell q q'/\nu q q'$. The masses of the first- and second-generation squarks are assumed to be 3~\tev, while all other sparticles, including third-generation squarks, are set above 5~\tev.
\begin{figure}[h]
\centering
\includegraphics[width=0.24\textwidth]{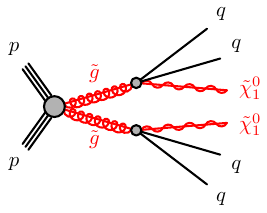}
\includegraphics[width=0.24\textwidth]{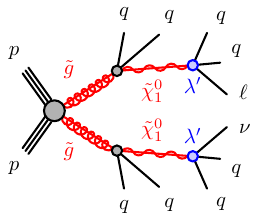}
\includegraphics[width=0.24\textwidth]{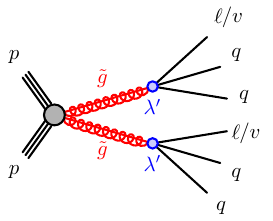}
\caption{Production and decay processes for the Gqq+LQD model. The dominant process varies with increasing $\lambda'$ coupling from left to right.}
\label{fig:diagrams1p1}
\end{figure}

\item[Gqq+UDD model:] This model contains light gluinos and a 200~GeV bino-like $\ninoone$ LSP, with non-zero $\lambdapp_{112}$ coupling and all other RPV couplings set to zero.
The gluinos are pair-produced and decay via off-shell squarks of the first and second generation.
As shown in Figure \ref{fig:diagrams1p4}, for low values of the RPV coupling, the gluino decays as $\gluino \rightarrow q q \ninoone$ \mbox{($q$ = $u$, $d$, $s$, $c$)} with the subsequent LSP decay, $\ninoone \rightarrow qqq$.
For larger values of the coupling, the gluino can also decay as $\gluino \rightarrow qqq$.
The masses of the first and second generation squarks are assumed to be 3~\tev, while the masses of the other sparticles are set above 5~\tev, including third generation squarks.

\begin{figure}[h]
\centering
\includegraphics[width=0.24\textwidth]{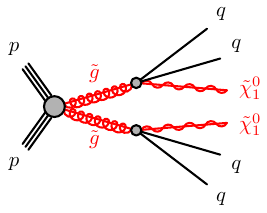}
\includegraphics[width=0.24\textwidth]{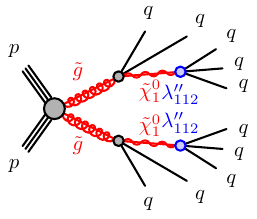}
\includegraphics[width=0.24\textwidth]{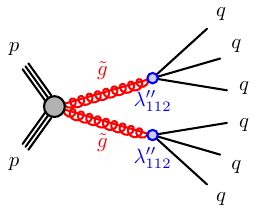}
\caption{Production and decay processes for the Gqq+UDD model. The dominant process varies with increasing \lambdapp coupling from left to right.}
\label{fig:diagrams1p4}
\end{figure}

\item[Gtt model:] This model contains light gluinos and a 200~GeV bino-like $\ninoone$ LSP, with non-zero $\lambdapp_{323}$ coupling and all other RPV couplings set to zero.
The gluinos are pair-produced and decay via off-shell top squarks.
As shown in Figure \ref{fig:diagrams1p2}, for low values of the RPV coupling, the gluino decays as $\gluino \rightarrow t t \ninoone$ with the subsequent LSP decay, $\ninoone \rightarrow tbs$.
For larger values of the coupling, the gluino can also decay as $\gluino \rightarrow t b s$.
The masses of the third-generation squarks are assumed to be 3 \tev; the masses of the other sparticles are set above 5 \tev, including first- and second-generation squarks.
\begin{figure}[h]
\centering
\includegraphics[width=0.24\textwidth]{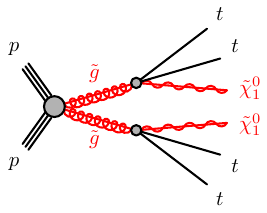}
\includegraphics[width=0.24\textwidth]{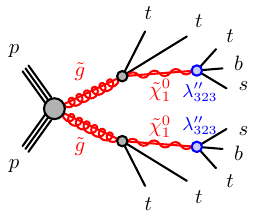}
\includegraphics[width=0.24\textwidth]{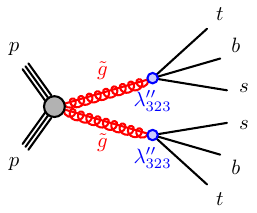}
\caption{Production and decay processes for the Gtt model. The dominant process varies with increasing \lambdapp coupling from left to right.}
\label{fig:diagrams1p2}
\end{figure}

\item[Stop model:] This model contains a light top squark, \stopone, and a 200~GeV bino-like $\ninoone$ LSP, with non-zero $\lambdapp_{323}$ coupling and all other RPV couplings set to zero.
As shown in Figure \ref{fig:diagrams1p3}, top squarks are pair-produced and decay as $\stopone \rightarrow t \ninoone$ for low coupling values, or $\stopone \rightarrow b s$ for high coupling strengths.
The LSP always decays as $\ninoone \rightarrow t b s$. The masses of the other sparticles are set above 3 \tev, but since the RPV decay proceeds via the right-handed top squark, which is already part of the simplified model, there is no impact on the lifetime or the branching fractions from this choice.
For high coupling values, single top-squark resonant production is also considered, $p p \rightarrow \stopone \rightarrow b s$, as shown in the rightmost diagram of Figure~\ref{fig:diagrams1p3}, leading to a di-jet final state that may provide stronger constraints on the top-squark mass than pair production at the LHC~\cite{Monteux:2016gag}. Single top-squark resonant production followed by a decay to a top quark and neutralino,  $p p \rightarrow \stopone \rightarrow t \ninoone$, is not considered because the branching fraction $B(\stopone \rightarrow bs)$ is dominant.
The cross section for single production is more than two orders of magnitude higher than pair production for a top-squark mass of 500 \GeV\ with $\lambdapp_{323}=1$, and evolves as $(\lambdapp_{323})^2$.
\begin{figure}[h]
\centering
\includegraphics[width=0.24\textwidth]{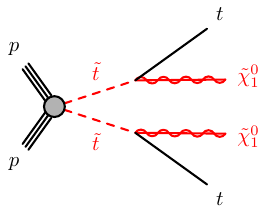}
\includegraphics[width=0.24\textwidth]{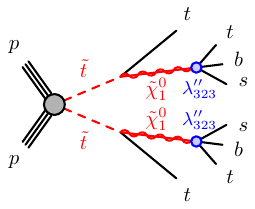}
\includegraphics[width=0.24\textwidth]{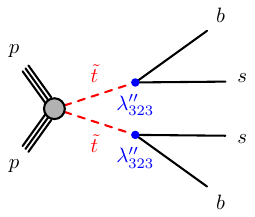}
\raisebox{0.4\height}{\includegraphics[width=0.24\textwidth]{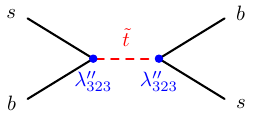}} \\
\caption{Production and decay processes for the stop model. The dominant process varies with increasing \lambdapp coupling from left to right. The single top-squark resonant production becomes dominant for high \lambdapp coupling values.}
\label{fig:diagrams1p3}
\end{figure}

\item[Stau model:] This model contains a light $\tau$-slepton ($\tau$-sneutrino), \stau~($\snu_\tau$), and a 50~GeV bino-like $\ninoone$ LSP, with a non-zero $\lambda_{i33}$ ($i \in \{1,2\}$) coupling and all other RPV couplings set to zero.
As shown in Figure~\ref{fig:diagrams2p2}, the $\tau$-slepton ($\tau$-sneutrino) are pair-produced and decay as $\stau \rightarrow \tau \ninoone$ ($\snu_\tau \rightarrow \nu \ninoone$) for small RPV couplings, or $\stau \rightarrow \ell\nu/\tau\nu$ ($\snu_\tau \rightarrow \tau \ell$), where $\ell=e,\mu$, for high coupling values. The LSP always decays as $\ninoone \rightarrow \tau\ell\nu/\tau\tau\nu$. The masses of the other sparticles are set above 5 \tev.
The pair-production of both left-handed and right-handed \stau~($\snu_\tau$) with the same mass are considered.
For a left-handed $\tau$-slepton, $B(\tilde{\tau} \rightarrow \ell \nu )$ is always zero, while for right-handed $\tau$-sleptons $B(\tilde{\tau} \rightarrow \tau \nu )=B(\tilde{\tau} \rightarrow \ell \nu )$.
For the bino-like \ninoone, independently of the $\lambda_{i33}$ coupling values, $B( \tilde{\chi}_1^0 \rightarrow \ell\tau\nu)=B( \tilde{\chi}_1^0 \rightarrow \tau\tau\nu)$.

\begin{figure}[h]
\centering
\includegraphics[height=0.15\textheight, width=0.24\textwidth]{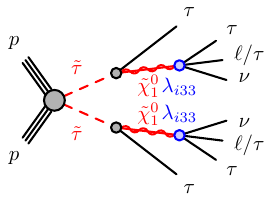}
\includegraphics[height=0.15\textheight, width=0.24\textwidth]{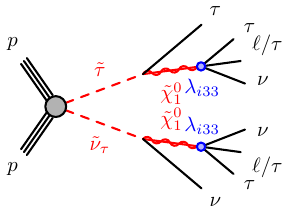}
\includegraphics[height=0.15\textheight, width=0.24\textwidth]{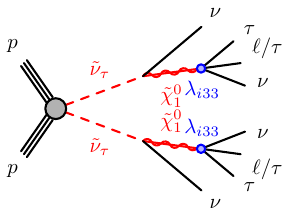}\\
\includegraphics[height=0.15\textheight, width=0.24\textwidth]{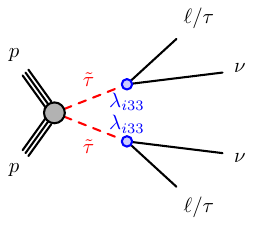}
\includegraphics[height=0.15\textheight, width=0.24\textwidth]{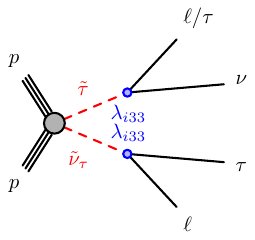}
\includegraphics[height=0.15\textheight, width=0.24\textwidth]{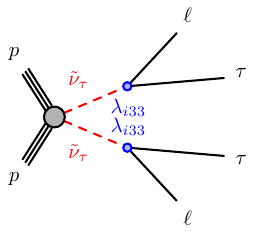}

\caption{Production and decay processes for the stau model. The top shows the dominant process of $\tilde{\tau}$ and $\tilde{\nu}$ decay into $\ninoone$ and $\tau$ or $\nu$ for relatively small values of $\lambda_{i33}$, while the bottom shows that of $\tilde{\tau}$ and $\tilde{\nu}$ decay into SM particles for relatively large values of $\lambda_{i33}$, where $i \in \{1,2\}$. }
\label{fig:diagrams2p2}
\end{figure}

\item[Higgsino model:] This model assumes nearly mass-degenerate higgsino-like $\tilde{\chi}_1^0$, $\tilde{\chi}_1^{\pm}$, and $\tilde{\chi}_2^0$, with $\Delta m(\tilde{\chi}_1^{\pm},\ninoone)=0.25~\GeV$ and $\Delta m(\ninotwo, \ninoone)=0.5~\GeV$. The $\lambda_{i33}$ ($i \in \{1,2\}$) coupling is non-zero and all other RPV couplings are set to zero.
The production of $\tilde{\chi}_1^0 \tilde{\chi}_2^{0}$, $\tilde{\chi}_1^{\pm} \tilde{\chi}_1^{\pm}$, $\tilde{\chi}_1^0 \tilde{\chi}_1^{\pm}$, and $\tilde{\chi}_2^0 \tilde{\chi}_1^{\pm}$ are considered separately. As shown in Figure~\ref{fig:diagrams2p1}, for low values of the RPV coupling, $\tilde{\chi}_2^{0}$ and $\tilde{\chi}_1^{\pm}$ decays with the subsequent LSP decay, while for larger values of the coupling, $\tilde{\chi}_2^{0}$ and $\tilde{\chi}_1^{\pm}$ decays promptly via the RPV coupling.
The mass of $\tilde{\tau}$ and $\tilde{\nu}_{\tau}$ masses are fixed to 3 \TeV, while all other sparticle masses are set above 5 \TeV~to ensure their decoupling.
In the model considered, regardless of the magnitude of the $\lambda_{i33}$ couplings, the ratio of branching fractions $B(\ninoone \rightarrow \ell\tau\nu)/B(\ninoone \rightarrow \tau\tau\nu)$ is set to 0.5,
whereas $ B(\tilde{\chi}_1^{\pm} \rightarrow \tau \nu \nu ) = B(\tilde{\chi}_1^{\pm} \rightarrow \ell \nu \nu ) = 0.5 \times B(\tilde{\chi}_1^{\pm} \rightarrow \tau \tau \ell ) $.
\begin{figure}[h]
\centering
\includegraphics[height=0.15\textheight, width=0.24\textwidth]{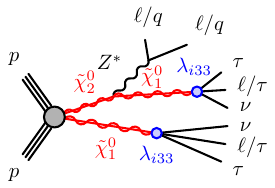}
\includegraphics[height=0.15\textheight, width=0.24\textwidth]{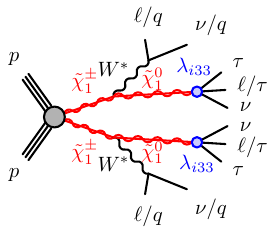}
\includegraphics[height=0.15\textheight, width=0.24\textwidth]{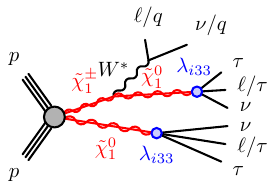}
\includegraphics[height=0.15\textheight, width=0.24\textwidth]{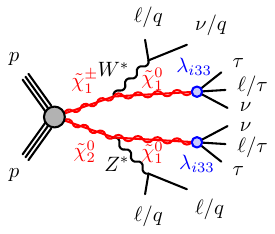} \\
\includegraphics[height=0.15\textheight, width=0.24\textwidth]{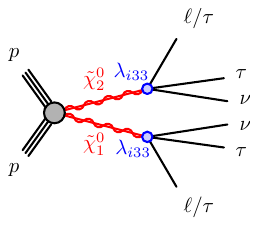}
\includegraphics[height=0.15\textheight, width=0.24\textwidth]{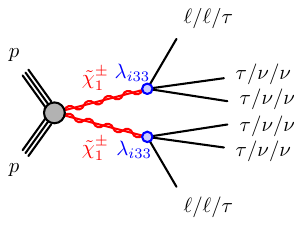}
\includegraphics[height=0.15\textheight, width=0.24\textwidth]{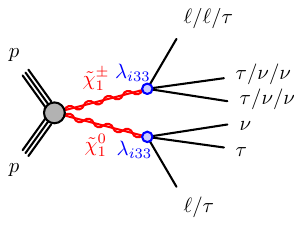}
\includegraphics[height=0.15\textheight, width=0.24\textwidth]{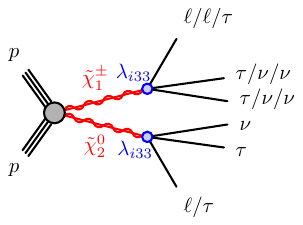} \\
\caption{Production and decay processes for the higgsino model. The top shows the dominant process of $\tilde{\chi}_2^0$ and $\tilde{\chi}_1^{\pm}$ decay into $\ninoone$ and off-shell \Zboson or \Wboson bosons for relatively small values of $\lambda_{i33}$, while the bottom shows that of $\tilde{\chi}_2^0$ and $\tilde{\chi}_1^{\pm}$ decay into SM particles for relatively large values of $\lambda_{i33}$, where $i \in \{1,2\}$.}
\label{fig:diagrams2p1}
\end{figure}

\end{description}

\subsection{External constraints}

The RPV couplings introduced in the previous section are constrained by theoretical requirements and experimental bounds from indirect decay measurements.

Theoretical upper limits on the $\lambdapp$ coupling strengths are obtained by considering the renormalization group equations of the superpotential parameters and by requiring the couplings to be perturbative up to the unification (GUT) scale, $\lambdapp(M_{GUT}) < \sqrt{4\pi}$.
The limits obtained are $\lambdapp_{112} < 1.25$ for the choice of parameters of Ref.~\cite{GOITY199569} and $\lambdapp_{323} < 1.07$ for those of Ref.~\cite{Allanach:1999mh}.
More stringent experimental limits on $\lambdapp_{112}$ can be obtained for particular choices of the sparticle masses.
Low-energy measurements such as di-nucleon decay~\cite{Zwane:2015bra,PhysRevLett.112.131803} impose
\mbox{$\lambdapp_{112} \lesssim 5 \cdot 10^{-7} \left( \frac{m(\tilde{q}_{R} )}{1~\tev} \right)^2 \left( \frac{m(\tilde{g})}{1~\tev} \right)^{1/2}$}. Combined with Eq.~(\ref{eq:lifetime}), the LSP lifetimes shorter than 100 ns are excluded for gluino masses up to 5 \tev, assuming a neutralino mass of 200 \gev.
The ATLAS analyses described in this paper that provide limits on $\lambdapp_{112}$ do not rely on $b$-quark identification, therefore the limits can be interpreted for any $\lambdapp_{ijk}$ coupling provided that $i,j,k \neq 3$.
The existing upper limit on the $\lambda_{133}$ coupling is derived from the experimental limit on the electron neutrino mass of the KATRIN experiment \cite{KATRIN:2021uub}.
The $\lambda_{133}$ term provides a Majorana mass via diagrams involving $\tau \tilde{\tau}$ loops \cite{Godbole:1992fb} to $\nu_e$.
By using the experimental limit on the electron neutrino mass at 90\% confidence level, the upper limit of $\lambda_{133}$ is expressed as:
\begin{equation}
\lambda_{133} < 0.0014 \sqrt{\frac{m(\tilde{\tau})}{100\ \GeV}}.
\end{equation}

The upper limit of the $\lambda_{233}$ coupling is calculated by considering the branching fraction ratio $R_{\tau}$, which is defined as:
\begin{equation}
R_{\tau} = \Gamma(\tau \rightarrow e \nu \bar{\nu}) / \Gamma(\tau \rightarrow \mu \nu \bar{\nu}).
\end{equation}
By comparing the SM prediction \cite{ledroit:in2p3-00362621} and one standard deviation ($1\sigma$) from the fitting result of PDG \cite{Workman:2022ynf}, which is $ 1/R_{\tau} = (97.62 \pm 0.28) \times 10^{-2} $,
the final upper limit of the $\lambda_{233}$ coupling is:
\begin{equation}
\lambda_{233} < 0.047 \frac{m(\tilde{\tau})}{100\ \GeV}.
\end{equation}

The external upper limit on the $\lambdap_{111}$ coupling is subject to stringent constraints from various experimental observations \cite{Baltz:1997ar,Pas:2015eia}. One source of these constraints is obtained from neutrinoless double beta decay ($0\nu\beta\beta$), upper limits on $\lambdap_{111}$ can be expressed as~\cite{Allanach:1999ic} :
\begin{equation}
\lambdap_{111} \lesssim 5.2 \times 10^{-4} \left( \frac{m(\tilde{e} )}{100~\gev} \right)^2 \left( \frac{m(\tilde{\chi}^{0})}{100~\gev} \right)^{1/2}
\end{equation}

\subsection{Signal sample production}
\label{subsec:sampleproduction}

Samples of Monte Carlo (MC) simulated events are used to model the signal.
The response of the detector to signal events is modeled with the full ATLAS detector simulation~\cite{SOFT-2010-01, Lukas:1458503} based on {\scshape Geant4}~\cite{Agostinelli:2002hh}.
The effect of multiple interactions in the same and neighbouring bunch crossings (\pileup) was modelled by overlaying the simulated hard-scattering event with inelastic proton--proton events generated with \PYTHIA[8.186]~\cite{Sjostrand:2007gs} using the \NNPDF[2.3lo] set of parton distribution functions (PDF)~\cite{Ball:2012cx} and the A3 set of tuned parameters~\cite{ATL-PHYS-PUB-2016-017}.
Signal samples are generated at leading order with \MGMCatNLOV 2.3.3~\cite{Alwall:2014hca} interfaced to \PYTHIA 8.210 with the A14 set of tuned parameters~\cite{ATL-PHYS-PUB-2014-021} for the modelling of the parton showering (PS) \cite{Mrenna:2016sih}, hadronisation and underlying event, with up to two additional partons in the matrix element (ME).
The ME–PS merging was carried out using the CKKW-L prescription \cite{Lonnblad:2001iq,Lonnblad:2011xx} with a matching scale set to one-quarter of the mass of the pair-produced SUSY particles.
The parton luminosities are provided by the {\scshape NNPDF23LO}~\cite{Ball:2012cx} PDF set.

Signal pair-production cross-sections are calculated to next-to-leading order (NLO) in the strong coupling constant, adding the resummation of soft-gluon emission at next-to-leading-logarithmic accuracy
(NLO+NLL) \cite{Beenakker:1996ch,Kulesza:2008jb,Kulesza:2009kq,Beenakker:2009ha,Beenakker:2011fu}. The nominal cross-section and its uncertainty are taken from an envelope of cross-section
predictions using different PDF sets as well as different factorization and renormalization scales, as described in Ref.~\cite{Borschensky:2014cia}. Although the models in this paper specify the squark masses, contributions from squarks are not considered in the gluino pair-production cross-sections.
The cross-section for single top-squark resonant production is computed to NLO in the strong coupling constant~\cite{Plehn:2000be,Monteux:2016gag}.
The cross-section for $\tau$-slepton and higgsino pair productions are calculated with \textsc{Resummino} v2.0.1 to NLO in the strong coupling constant~\cite{Bozzi:2007qr,Fuks:2013vua,Fiaschi:2018xdm,Fuks:2013lya,Fuks:2012qx,Fuks:2013vua,Debove:2010kf,Fiaschi:2018hgm}.

\begin{figure}[t]
\centering
\subfloat[]{\includegraphics[width=0.45\textwidth]{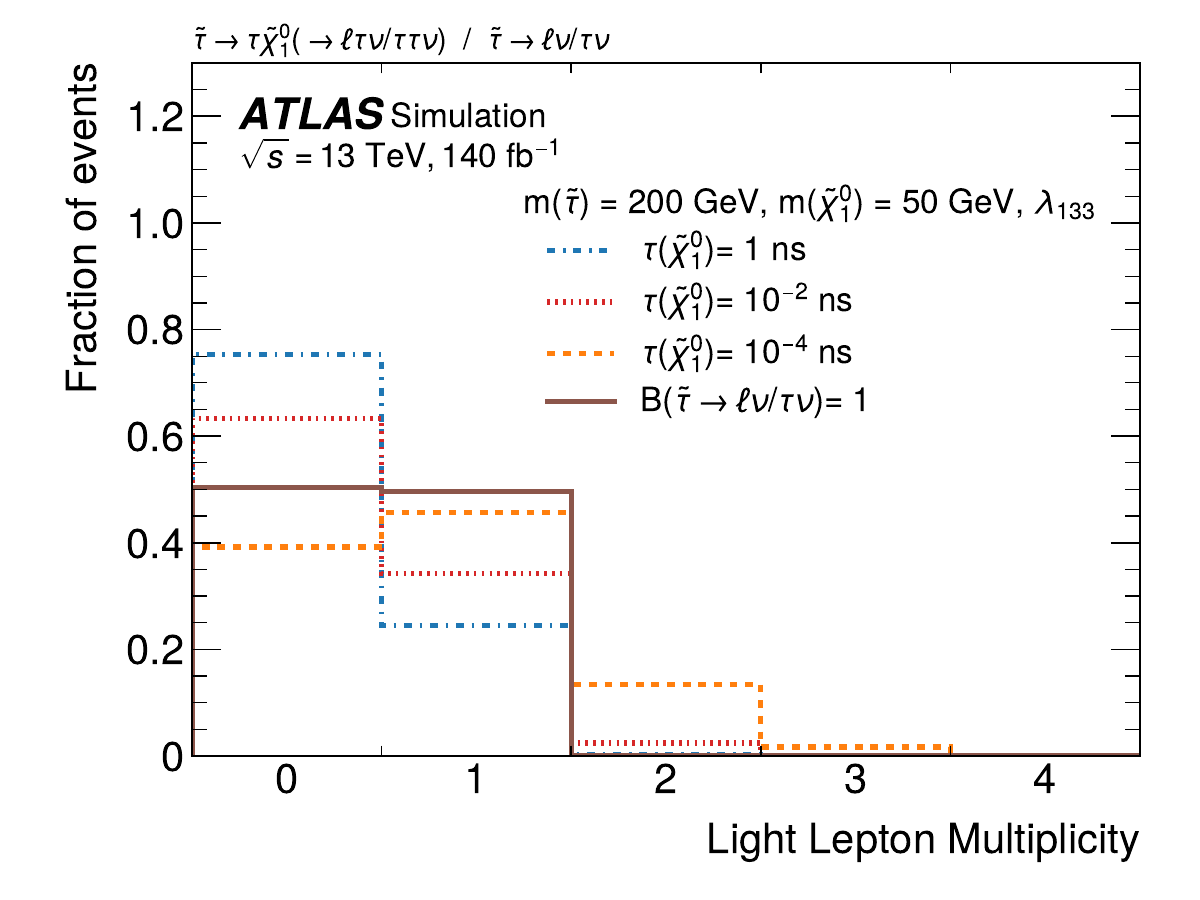}\label{fig:variables_a}}
\subfloat[]{\includegraphics[width=0.45\textwidth]{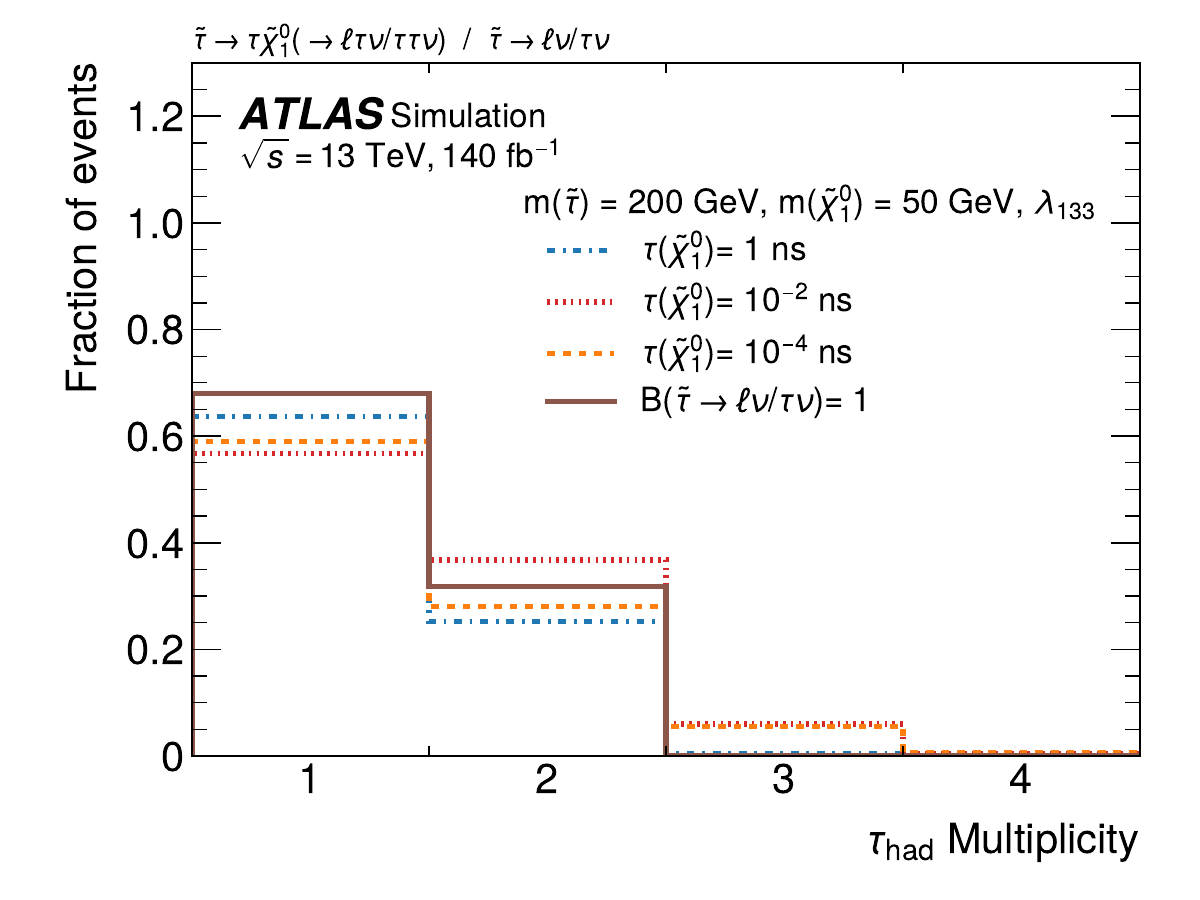}\label{fig:variables_b}} \\
\subfloat[]{\includegraphics[width=0.45\textwidth]{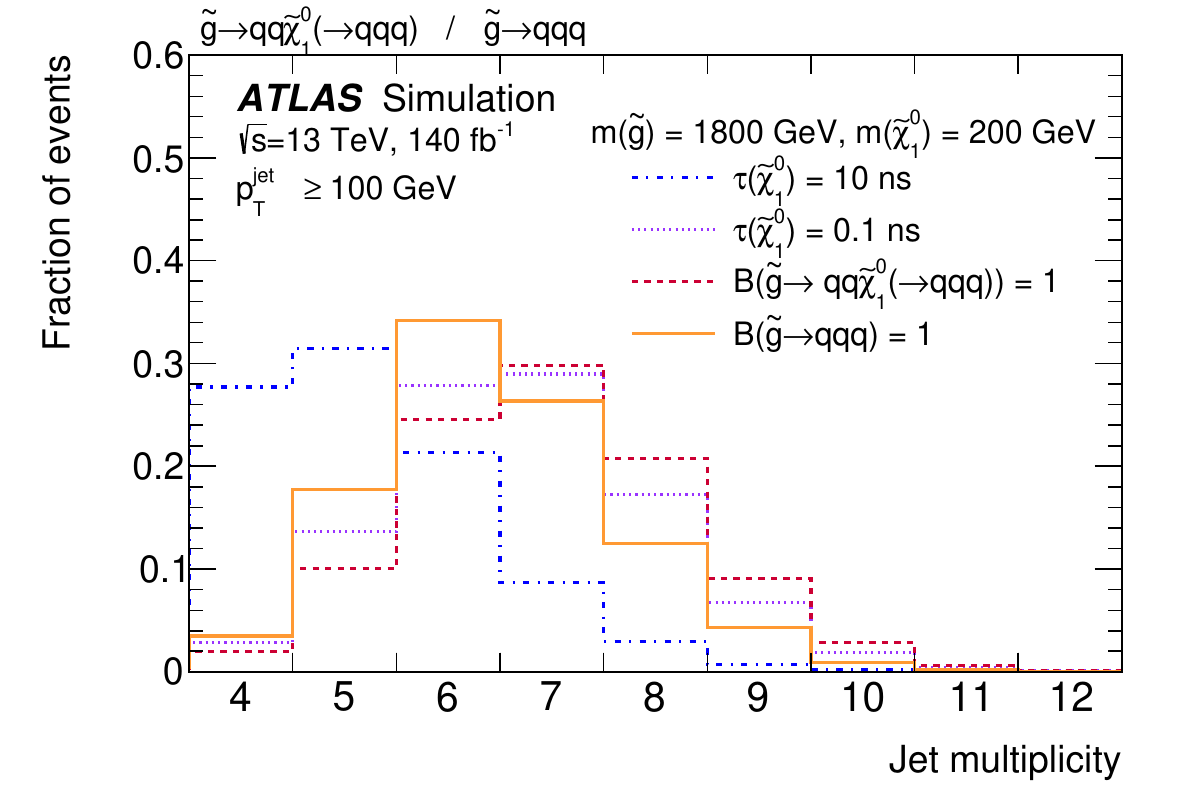}\label{fig:variables_c}}
\subfloat[]{\includegraphics[width=0.45\textwidth]{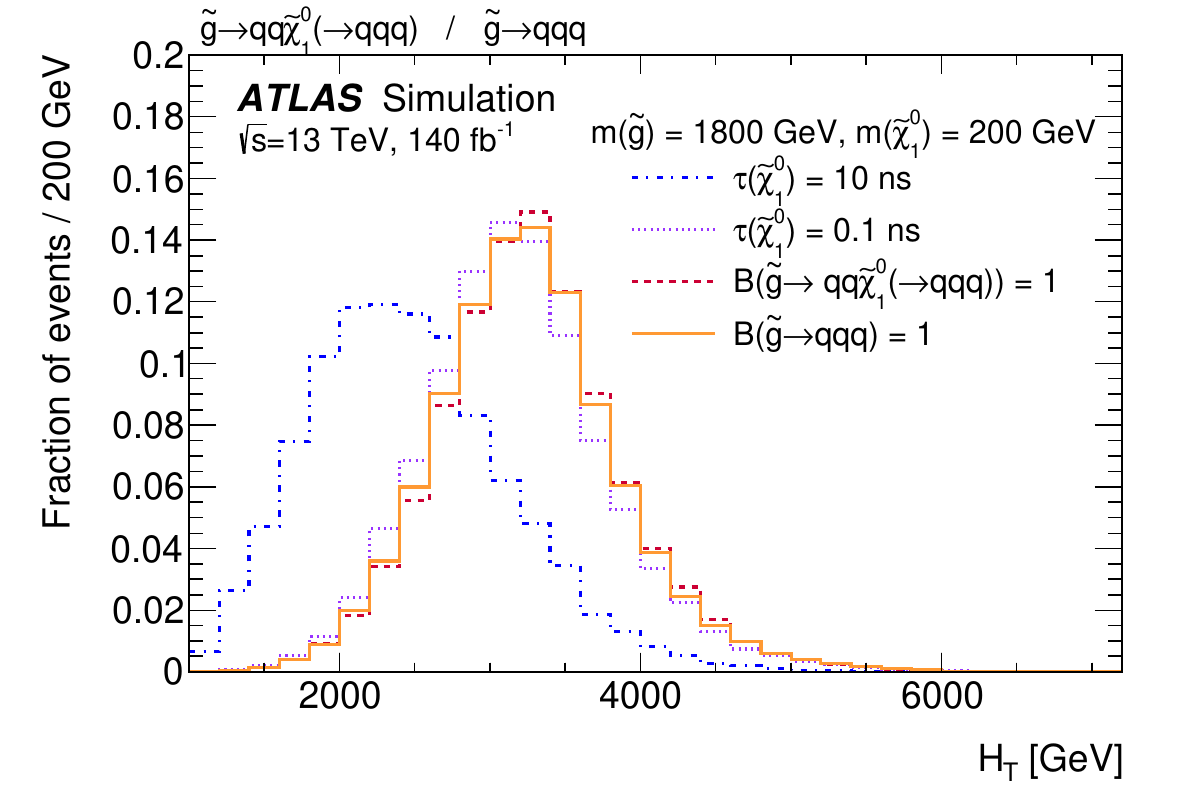}\label{fig:variables_d}} \\
\subfloat[]{\includegraphics[width=0.45\textwidth]{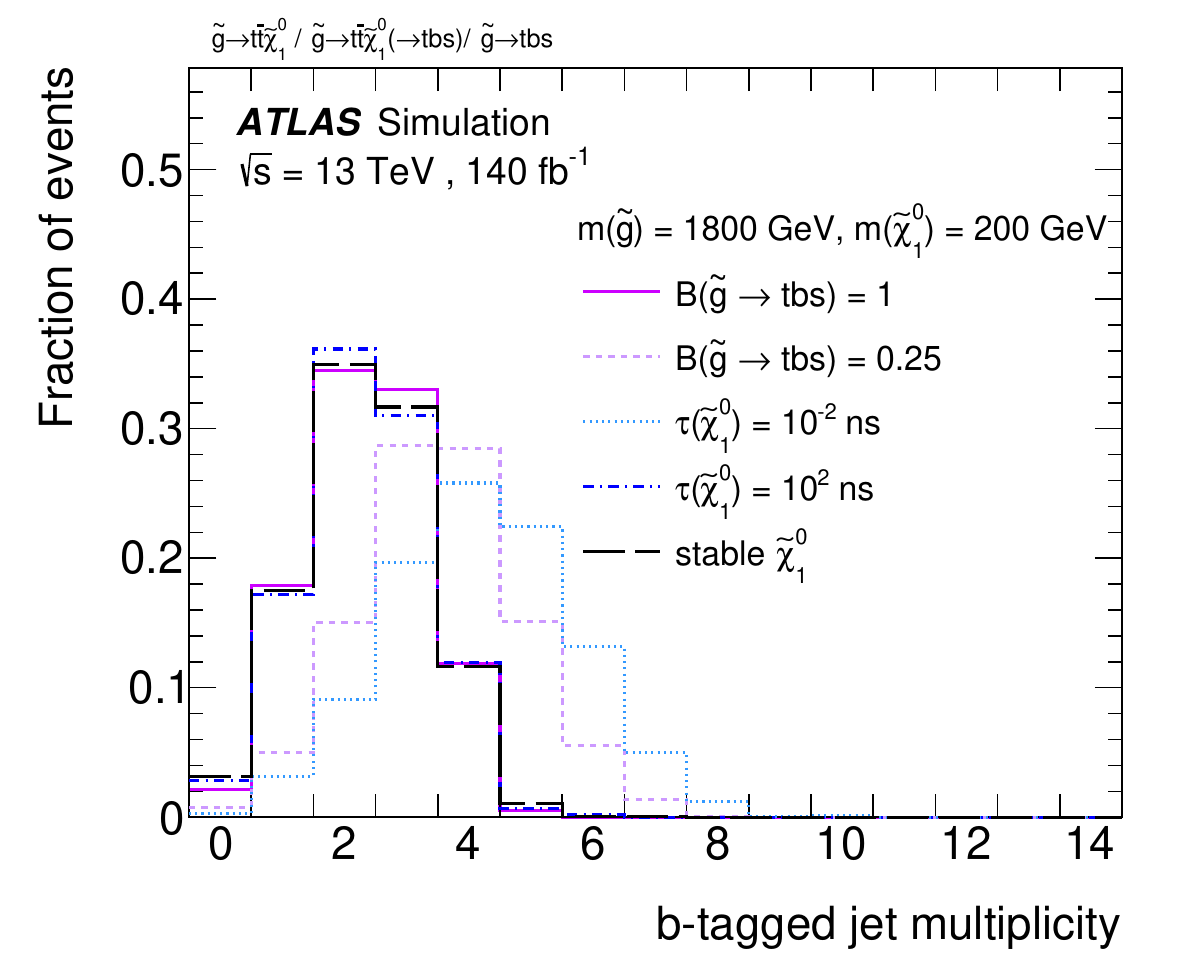}\label{fig:variables_e}}
\subfloat[]{\includegraphics[width=0.45\textwidth]{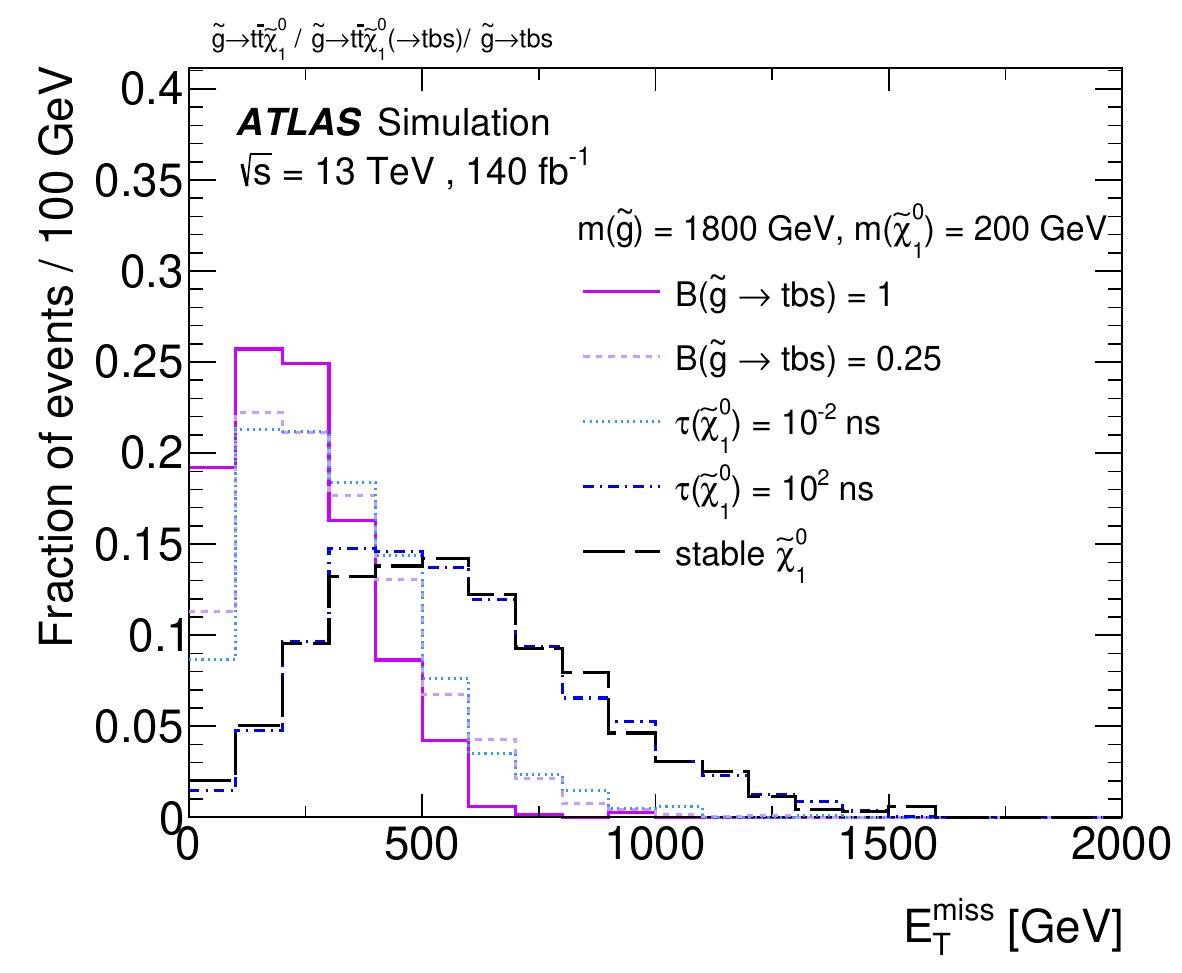}\label{fig:variables_f}}
\caption{Impact of neutralino decays with different lifetimes on the \protect\subref{fig:variables_a} number of light leptons ($\ell$), \protect\subref{fig:variables_b} number of hadronically-decaying $\tau$-leptons (\tauhad), \protect\subref{fig:variables_c} number of jets, \protect\subref{fig:variables_d} the $H_\mathrm{T}$, \protect\subref{fig:variables_e} number of $b$-jets and \protect\subref{fig:variables_f} \met.
All observables are shown at the reconstruction level with full simulation of the ATLAS detector. In (a) and (b), the stau model with non-zero $\lambda_{133}$ is shown. If the branching fraction (B) is specified, the gluino and $\tau$-slepton can decay directly to SM particles or via a prompt LSP decay. In (c) and (d) the Gqq+UDD model with non-zero $\lambdapp_{112}$ is shown. In (e) and (f), the Gtt model with non-zero $\lambdapp_{323}$ is shown.}

\label{fig:variables}
\end{figure}

Figure~\ref{fig:variables} shows the impact of neutralino decays with different lifetimes on six representative distributions of reconstructed-object multiplicities and several kinematic variables defined in Table~\ref{tab:variablelist}.
The benchmark models used for this comparison are selected for their representative topologies, specifically targeting final states with the highest multiplicities of jets, $b$-jets, and hadronically decaying $\tau$-leptons (\tauhad).
The multiplicity of reconstructed light leptons and \tauhad in the stau model (Figures~\ref{fig:variables_a} and~\ref{fig:variables_b}) have a strong dependency on the LSP lifetime ($\tau_\text{LSP}$).
The number of reconstructed light leptons increases once the neutralino decays are sufficiently prompt to be reconstructed in the detector.
In contrast, the \tauhad reconstruction is more robust against small displacements, as the reconstruction algorithm makes use of secondary vertex information~\cite{ATL-PHYS-PUB-2019-033}.
Similarly, the number of reconstructed jets increases once the neutralino decays are sufficiently prompt to be reconstructed in the calorimeter, as shown in Figure~\ref{fig:variables_c}.
A similar behaviour is also observed in the distribution of the hadronic transverse energy ($H_\mathrm{T}$), shown in Figure~\ref{fig:variables_d}, which increases with the number of jets that are reconstructed when the neutralino has a small lifetime.
The number of reconstructed $b$-jets, shown in Figure~\ref{fig:variables_e}, increases as the neutralino lifetime decreases since the $\ninoone$ decay is now sufficiently prompt to occur inside the detector volume, allowing for the reconstruction and identification of $b$-jets.
The number of $b$-jets drops once the $\ninoone$ decays promptly and enters the direct $\gluino$ decay regime since the absence of $\gluino \rightarrow tt \ninoone(\rightarrow tbs)$ reduces the production of $b$-jets.
The magnitude of the missing transverse momentum (\met), shown in Figure~\ref{fig:variables_f}, decreases for shorter lifetimes as a larger fraction of decays happen within the detector volume; the reduced values of \met strongly weaken the sensitivity of RPC searches.

\begin{table}[hptb]
\begin{center}
\caption{Summary of variable definitions in reinterpreted searches that are used in Figure~\ref{fig:variables} and to define the signal region selections described in Section~\ref{sec:analyses}.
}
\label{tab:variablelist}
\footnotesize{\renewcommand{\arraystretch}{1.3}
\makebox[\textwidth]{%
\begin{tabular}{|c|c |}
\hline
\hline
Variable	& Definition \\
\hline
$\mathbf{p}_\text{T}^{\text{miss}}$ & Negative vector sum of the $\mathbf{p}_\text{T}$ of all selected and calibrated physics objects in the event \\
\hline
\met & Magnitude of $\mathbf{p}_\text{T}^{\text{miss}}$ \\
\hline

$C$ &  Event-shape variable derived from the linearized sphericity tensor of the event~\cite{STDM-2019-02}  \\
\hline

$m_{\mathrm{eff}}$ & Scalar sum of lepton $p_\text{T}$, jet $p_\text{T}$ and \met \\
\hline
$H_\text{T}$ & Scalar sum of the transverse momenta of all jets with $\pt>50$~GeV and $|\eta|<2.8$ \cite{SUSY-2018-22} \\
\hline
\rule{0pt}{4ex}

$R_{DV}$  &  $R_{DV} = \sqrt{x_{DV}^2+y_{DV}^2}$, where $x_{DV}$, $y_{DV}$ are the Cartesian coordinates of the displaced vertex~\cite{SUSY-2018-13} \\
\hline

$\sum \ptjet$ & Scalar sum of the transverse momenta of all jets\\
\hline

$m_1^{R=1.2}$  &  Mass of leading jet which has been reclustered with radius parameter $R=1.2$~\cite{SUSY-2018-12}	 \\
\hline
\rule{0pt}{4ex}

$\mathcal{A}$ & Mass asymmetry, defined as $\mathcal{A}=\frac{|m_1-m_2|}{m_1+m_2}$, where $m_1$ and $m_2$ are the invariant masses of the \\
& two reconstructed dijet pairs~\cite{SUSY-2016-09}	  \\
\hline

$m_{jj}$   &  Invariant mass of dijet system \\
\hline

$m_{\mathrm{T2}}$ & ``stransverse mass", which has a kinematic endpoint for events where two massive pair \\ & produced particles each decay into a detected object and an undetected object~\cite{Lester:1999tx, Barr:2003rg} \\
\hline

$m_{\mathrm{Tsum}}$ & Sum of the transverse mass values of the two highest-$\pt$ $\tau$ candidates with the $\mathbf{p}_\text{T}^{\text{miss}}$~\cite{SUSY-2023-03} \\
\hline
\rule{0pt}{4ex}

$\Delta R_{\ell\ell}$ &  Opening angle between the two leptons, i.e. $\Delta R_{\ell\ell} = \sqrt{(\Delta\eta_{\ell\ell})^2+(\Delta\phi_{\ell\ell})^2}$~\cite{SUSY-2018-02} \\
\hline
\hline
\end{tabular}
}
}
\end{center}
\end{table}
\FloatBarrier


%
\section{Analysis reinterpretation}
\label{sec:analyses}


A set of thirteen ATLAS searches that are sensitive to the models described in Section~\ref{sec:susymodels} are reinterpreted.
None of the analyses observed a significant excess above the SM expectation in 140~fb$^{-1}$ of 13~\tev\ proton--proton collision data; consequently, they are used to set 95\% confidence level exclusion limits.
An outline of each of the included analyses is presented below.

The MC signal samples described in Section~\ref{subsec:sampleproduction} are used as input for either the \texttt{RECAST} framework or the original analysis code.
The \texttt{RECAST} framework preserves the full analysis workflow, archiving event yields from background MC samples and observed data, and storing them in Docker containers~\cite{10.5555/2600239.2600241}.
This enables the full re-execution of the event selection, systematic uncertainty evaluation, and statistical analysis for new signal samples.
When no additional systematic uncertainties are required,
the final exclusion limits are set by CLs~\cite{Read_CLs} values obtained either directly from the \texttt{RECAST} implementation (typically using \textsc{HistFitter}~\cite{Baak:2014wma}) or by performing a statistical analysis with \textsc{pyhf}~\cite{pyhf_zenodo, pyhf_joss} on the \texttt{RECAST} output.
If additional systematic uncertainties are included, as described in Section~\ref{sec:obj_and_uncert}, the workflow is adapted: for \texttt{RECAST}, the output signal likelihood is modified to incorporate these uncertainties; for the original analysis code, the additional systematics are added to the signal expectation during the limit-setting.
The final statistical analysis is subsequently performed using \textsc{HistFitter} or \textsc{pyhf}.
In both cases, the preservation of the exact background likelihood from the original analysis is guaranteed.

\begin{description}

\item[RPV multijet:]
this analysis \cite{SUSY-2019-24} searches for gluino pair production with subsequent RPV decays into quarks in events with a large number of jets. Events with energetic charged leptons are explicitly excluded, and there are no requirements on missing transverse momentum. %
This analysis has sensitivity to the Gqq+UDD model, shown in Figure~\ref{fig:diagrams1p4}.

\item[RPC 0-lepton (0L), 2-6 jets:]
this analysis~\cite{SUSY-2018-22} searches for pair production of squarks or gluinos in final states with jets and \met, while vetoing electrons and muons. Three strategies are used: a ‘multi-bin search’ in which the background expectations are fit to the observed data yields in multiple event selection bins simultaneously, a machine learning based selection using boosted decision trees, and a simple single-bin cut-and-count approach giving sensitivity to generic models characterized by the above final states. %
This analysis has sensitivity to the Gqq+LQD and Gqq+UDD models, shown in Figures~\ref{fig:diagrams1p1} and \ref{fig:diagrams1p4}, respectively, and is reinterpreted with the \texttt{RECAST} framework.

\item[RPV DV+jets:]
this analysis~\cite{SUSY-2018-13} searches for decays of new, long-lived particles using events with a massive, multi-track displaced vertex and multiple energetic jets. The dedicated reconstruction techniques employed in the analysis significantly increase the sensitivity to long-lived particles decaying in the inner detector. Background estimates for SM processes are derived from data. The results are interpreted with two production scenarios: the associated production of the lightest charged and neutral electroweakinos, $\tilde{\chi}_1^{\pm}\tilde{\chi}_1^{0}$, $\tilde{\chi}_1^{+}\tilde{\chi}_1^{-}$, $\tilde{\chi}_2^{0}\tilde{\chi}_1^{0}$, $\tilde{\chi}_1^{\pm}\tilde{\chi}_2^{0}$, assuming pure-higgsino states, and the production of gluino pairs where each gluino decays promptly with a 100\% branching fraction to two quarks and a neutralino. The neutralino subsequently decays via the \lambdapp coupling to light-flavour quarks. %
This analysis has sensitivity to the Gqq+UDD, Gtt, and Stop models, as shown in Figures~\ref{fig:diagrams1p4}, \ref{fig:diagrams1p2}, and \ref{fig:diagrams1p3}, respectively.

\item[RPC multi-$b$:]
this analysis~\cite{SUSY-2018-30} targets gluino production with the subsequent decay to third-generation squarks and a neutralino, in a scenario equivalent to the Gtt model in the RPC limit.
The search requires high jet and $b$-jet multiplicity, zero or one lepton and large \met .
The sensitivity of the search is optimized using both a cut-and-count and a deep neural network approach.
This analysis has sensitivity to the Gtt model, as shown in Figure~\ref{fig:diagrams1p2}, and is reinterpreted with the \texttt{RECAST} framework.

\item[RPV same-sign and 3-leptons (SS/3L):]
this analysis~\cite{SUSY-2020-27} covers RPV gluino pair production models that share similar final states to the targeted scenarios of the Gtt model.
The requirement of two same-sign leptons provides a powerful handle to suppress the SM backgrounds,
and allows the search to design SRs without an \met\ requirement, in order to cover RPC and RPV scenarios. %
This analysis has sensitivity to the Gqq+LQD and Gtt models, as shown in Figure~\ref{fig:diagrams1p1} and \ref{fig:diagrams1p2}, respectively.

\item[RPV 1-lepton (1L):]
this analysis~\cite{SUSY-2019-04} searches for pair production of supersymmetric particles with RPV couplings, yielding final states with at least one lepton, very high jet multiplicity and either no $b$-jets or many $b$-jets. %
This analysis has sensitivity to the Gtt and Stop models, as shown in Figures~\ref{fig:diagrams1p2} and \ref{fig:diagrams1p3}, respectively, and is reinterpreted with the \texttt{RECAST} framework.

\item[RPC stop 0-lepton (0L):]
this analysis~\cite{SUSY-2018-12} searches for top-squark pair production in a \ttbar+\met\ final state, with both tops decaying hadronically. This analysis exploits jet reclustering to reconstruct boosted hadronic top decays, while requiring that other quantities are incompatible with SM top-quark pair production.
This analysis has sensitivity to the Stop model, shown in Figure~\ref{fig:diagrams1p3}, and is reinterpreted with the \texttt{RECAST} framework.

\item[RPV stop dijet:]
this analysis~\cite{SUSY-2016-09} targets top-squark pair production with the subsequent decay to a $b$-quark and $s$-quark, in a scenario equivalent to the stop model considered in this paper in the regime with very high RPV coupling. The analysis requires two pairs of jets with large mass and low mass asymmetry. %
This analysis has sensitivity to the Stop model, as shown in Figure~\ref{fig:diagrams1p3}.

\item[RPV dijet resonance:]
this analysis~\cite{EXOT-2019-03} searches for a new resonance decaying into two energetic jets
with an inclusive jet selection and with separate selections where at least one or exactly two
jets are $b$-jets. By including dedicated optimizations for the $b$\bbar, $bq$ and $bg$ final states, the sensitivity to the dijet resonance is significantly increased. This analysis has sensitivity to the Stop model, as shown in Figure~\ref{fig:diagrams1p3}.

\item[RPV multi-$b$:]
this analysis~\cite{SUSY-2018-38} searches for top-squark pair production in events with no leptons, no \met , a high multiplicity of energetic jets and a large number of $b$-tagged jets. %
A simple cut-and-count strategy based solely on the jet and $b$-tagged jet multiplicities is adopted.
This analysis has sensitivity to the Gtt model, as shown in Figure~\ref{fig:diagrams1p2}.

\item[RPC Di-$\tau$:]
this analysis~\cite{SUSY-2023-03} targets RPC SUSY models where $\tilde{\chi}_1^{\pm} \tilde{\chi}_2^{0}$ and $\tilde{\chi}_1^{\pm} \tilde{\chi}_1^{\mp}$ decay via a $\tilde{\tau}$ particle to a final state with at least two \tauhad and large \MET.
The analysis relies on two transverse-mass related variables, $m_{\mathrm{T2}}$ and $m_{\mathrm{Tsum}}$, defined in Table~\ref{tab:variablelist}, to discriminate signal against the SM background.
Given the high $\tau$-lepton multiplicity in events with moderate or large $\lambda_{i33}$ ($i \in \{1,2\}$) coupling, and the possibility to obtain large \MET~due to high momentum neutrinos or from long-lived \ninoone's that fail to be reconstructed, this search has sensitivity to the Stau and Higgsino models, as shown in Figures~\ref{fig:diagrams2p2} and~\ref{fig:diagrams2p1}, respectively.

\item[RPC and RPV (RPC/RPV) Four-lepton (4L):]
this analysis~\cite{SUSY-2018-02} targets prompt SUSY models with non-zero RPV $\lambda_{121/122}$ or $\lambda_{133/233}$ couplings, as well as models with RPC General Gauge Mediation~(GGM) scenarios that predict events characterized by large charged-lepton multiplicity in the final state.
The analysis relies on the \MET~for RPC GGM scenarios and the effective mass ($m_{\mathrm{eff}}$), defined in Table~\ref{tab:variablelist}, for RPV models to discriminate signal against the SM background.
Given the high lepton multiplicity in events with moderate or large $\lambda_{i33}$ ($i \in \{1,2\}$) coupling, and the possibility to obtain large \MET~due to high momentum neutrinos or from long-lived \ninoone that fail to be reconstructed because of the large $\tau_\text{LSP}$, this search has sensitivity to the Stau and Higgsino models, as shown in Figures~\ref{fig:diagrams2p2} and~\ref{fig:diagrams2p1}, respectively. This analysis is reinterpreted with the \texttt{RECAST} framework.

\item[RPC Displaced Lepton (DL):]
this analysis~\cite{SUSY-2018-14} searches for charged leptons with large impact parameters, which provides unique sensitivity to long-lived sleptons. Three orthogonal signal regions are defined which all require at least two light leptons and are distinguished by the flavor of the leading and sub-leading light lepton. %
This search has sensitivity to the Higgsino model, as shown in Figure~\ref{fig:diagrams2p1}, and is reinterpreted with the \texttt{RECAST} framework.

\end{description} %

The main characteristics of the most sensitive signal regions in each of the above analyses is summarized in Table~\ref{tab:SRsummary}. The requirements on \met or related variables, defined in Table~\ref{tab:variablelist}, are shown for each analysis, highlighting the different approach of RPC and RPV searches.
All the signal regions from the corresponding analyses are considered in the limit setting procedure, even if not listed in Table~\ref{tab:SRsummary}, except where explicitly noted.

\begin{table}[hptb]
\begin{center}
\caption{Main characteristics of the most sensitive signal region per analysis. Only an illustrative subset of the selections that define each signal region are included here. The definition of different kinematic variables used for these selections is described in Table~\ref{tab:variablelist}. The $\tauhad^{M}$ ($\tauhad^{T}$) refers to \tauhad passing the \texttt{Medium} (\texttt{Tight}) identification working points~\cite{ATL-PHYS-PUB-2019-033}.  A dash (--) is used to indicate that the variable is not used in the analysis selection. The requirement of two same-sign leptons is denoted as SS. The OS$_{1,2}$ means that the leading and second leading taus in terms of transverse momentum must have opposite sign. The definition of the different kinematic variables used for cuts indicated in the table can be found in the corresponding publications for the input analyses. The RECAST column indicates that the entry is a recast result, not the original analysis.
}
\label{tab:SRsummary}
\setlength{\tabcolsep}{0.15pc}
\footnotesize{\renewcommand{\arraystretch}{1.3}
\makebox[\textwidth]{%
\begin{tabular}{|l|c c c c c c c|}
\hline
\hline
\multirow{2}{*}{Analysis name} &  \multirow{2}{*}{$\ell/\tauhad^{M}/\tauhad^{T}$} &  \multirow{2}{*}{ Jets / $b$-tags}
& \multirow{2}{*}{\begin{minipage}{2 cm}  \met \newline \centering selection \end{minipage}}
& \multirow{2}{*}{\begin{minipage}{2 cm} Representative \newline \centering cuts \end{minipage}}
& \multirow{2}{*}{\begin{minipage}{1 cm} Model \newline \centering targeted \end{minipage}}  & \texttt{RECAST}& \\
&         & 	    &  &  &  &  & \\
\hline

RPV MJ \cite{SUSY-2019-24} 		& 0 / -- / --  		& $\geq 8$ / $\geq 2$  & --	& $C\geq0.85$ 		& Gqq+UDD & No &\\

\hline

\multirow{2}{*}{RPC 0L, 2-6 jets \cite{SUSY-2018-22}} 		& \multirow{2}{*}{0 / -- / --}  		& \multirow{2}{*}{$\geq 6$ / --}   & \multirow{2}{*}{$\met/\sqrt{H_{T}} > 10$}	& \multirow{2}{*}{$\meff > 3400$ \gev} 		& Gqq+LQD & \multirow{2}{*}{Yes}& \\
&&&&& Gqq+UDD   &&\\
\hline

\multirow{2}{*}{RPV DV+jets \cite{SUSY-2018-13}}		&	\multirow{2}{*}{--}  &	\multirow{2}{*}{$\geq 4$ / --}		&	\multirow{2}{*}{--}		&	\multirow{2}{*}{$R_{DV}<300$ mm}	& Gqq+UDD, & \multirow{2}{*}{No} & \\
&&&&& Gtt, Stop & &\\
\hline

\multirow{2}{*}{RPC multi-$b$ \cite{SUSY-2018-30} }			&	0 / -- / -- 	& $\geq 5$ / $\geq 3$		&	$\met > 600$ \gev		&	$\meff > 2900$ \gev		&	\multirow{2}{*}{Gtt} & \multirow{2}{*}{Yes} &\\
&	1 / -- / -- 				&	$\geq 4$ / $\geq 3$	& $\met > 300$ \gev                   & $\meff > 800$ \gev 					& & & \\
\hline
\multirow{3}{*}{\begin{minipage}{3.4cm}RPV SS/3L \cite{SUSY-2020-27}\end{minipage}}	& \multirow{3}{*}{2 SS  / -- / -- } & $\geq 5$ / -- &	\multirow{3}{*}{--}	& $\meff>2600$ \gev & \multirow{2}{*}{\begin{minipage}{2 cm} Gqq+LQD, \newline \centering Gtt \end{minipage}}  & \multirow{3}{*}{No} & \\
&					&		$\geq 6$ / $=1$	&							& $\sum \ptjet>1600$ \gev &  & &\\
&					&		$\geq 5$ / $\geq 2$	&							& $\sum \ptjet>1600$ \gev &  &   &\\
\hline

RPV 1L \cite{SUSY-2019-04}	&	1 / -- / -- 				&	$\geq 10$ / $\geq 3$	&	--				&	--					& Gtt, Stop & Yes &\\
\hline

RPC Stop 0L	\cite{SUSY-2018-12}		&	0 / -- / -- 			&	$\geq 4$ / $\geq 2$			&	$\met > 250$ \gev		&	$m_1^{R=1.2}>120$ \gev	& Stop & Yes &\\
\hline
RPV stop dijet \cite{SUSY-2016-09}		&	--  		&	$\geq 4$ / $\geq 2$			&	--		&	$\mathcal{A}<0.05$		& Stop & No &\\
\hline

RPV dijet resonance \cite{EXOT-2019-03}		&	--  		&	$\geq 2$ / $\geq 2$		&	--		&	$m_{jj}>1133$ GeV	& Stop & No &\\
\hline

{RPV multi-$b$ \cite{SUSY-2018-38} }			&	0 / -- / -- 	& $\geq 9$ / $\geq 5$	 &	-- &	--	&{Gtt} & No &\\
\hline

\multirow{6}{*}{RPC Di-$\tau$ \cite{SUSY-2023-03} }		& -- / $= 2$, OS$_{1,2}$~/ $\geq 1$  	& -- / veto  		& $\in [60,\ 150]\ \GeV$ 	& $ m_{\mathrm{T2}} > 80\ \GeV$	& & \multirow{6}{*}{No}  &\\
\cline{2-5}
& \multirow{2}{*}{-- / $= 2$, OS$_{1,2}$~/ --  }			& \multirow{2}{*}{-- / veto}  	& \multirow{2}{*}{$ > 150\ \GeV$} 	& $ m_{\mathrm{T2}} > 85\ \GeV$	& &   &\\
& 		&   &  	& $m_{\mathrm{Tsum}} > 400\ \GeV$	&  Stau, &  &\\
\cline{2-5}
& -- / $\geq 2$, OS$_{1,2}$~/ $\geq 1$  	& -- / veto  		& $\in [60,\ 150]\ \GeV$ 	& $ m_{\mathrm{T2}} > 70\ \GeV$	& Higgsino &  &\\
\cline{2-5}
& \multirow{2}{*}{-- / $\geq 2$, OS$_{1,2}$~/ --  }			& \multirow{2}{*}{-- / veto}  		& \multirow{2}{*}{$ > 150\ \GeV$} 	& $ m_{\mathrm{T2}} > 85\ \GeV$	& &   &\\
& 		&   &  		&  $m_{\mathrm{Tsum}} > 400\ \GeV$	& &   &\\
\hline
\multirow{8}{*}{RPC/RPV 4L \cite{SUSY-2018-02} }		& \multirow{4}{*}{$\geq 4$ / $\geq 0$ / --}  	& \multirow{4}{*}{-- / veto} 	& $ > 100\ \GeV$	& \multirow{2}{*}{--}	& & \multirow{8}{*}{Yes} &\\
& 		&   &  		 $ > 200\ \GeV$ 	& 	& &   &\\
\cline{4-5}
&  	&  & \multirow{2}{*}{ --}  		& $m_{\mathrm{eff}} > 600\ \GeV$ 	& &   &\\
& 		&   &  		& 	 $m_{\mathrm{eff}} > 1250\ \GeV$ 	&  Stau, &  &\\
\cline{2-5}
& \multirow{2}{*}{$= 3$ / $\geq 1$ / --}  	& \multirow{2}{*}{-- / veto}   	& \multirow{2}{*}{--}	& $m_{\mathrm{eff}} > 600\ \GeV$ 	&  Higgsino &  &\\
& 		&  &  		& 	 $m_{\mathrm{eff}} > 1000\ \GeV$ 	& &   &\\
\cline{2-5}
& \multirow{2}{*}{$= 2$ / $\geq 2$ / --}  	& \multirow{2}{*}{-- / veto}   	& \multirow{2}{*}{--}	& $m_{\mathrm{eff}} > 600\ \GeV$ 	& &  & \\
& 		&   &  		& 	 $m_{\mathrm{eff}} > 1000\ \GeV$ 	& &   &\\

\hline
{RPC DL~\cite{SUSY-2018-14} }	& 2 / -- / --  	& -- / -- 	&	-- & $\Delta R_{\ell\ell} > 0.2$	&{Higgsino} & Yes &\\
\hline
\hline
\end{tabular}
}
}
\end{center}
\end{table}

%
%


%
\section{Objects and systematic uncertainties}
\label{sec:obj_and_uncert}


Except for the DV+jets and DL analyses, the studies in this paper are not optimized for long-lived signatures and thus use the standard ATLAS reconstruction of prompt objects.
While the full ATLAS detector simulation based on {\scshape Geant4} provides a baseline estimate of reconstruction efficiency for displaced objects, dedicated systematic uncertainties are assigned to account for the degree to which the simulation accurately models these efficiencies as a function of displacement.
This section describes the systematic uncertainties introduced to the signal acceptance times efficiency from applying the standard algorithms to non-prompt objects.
The object definitions vary across analyses and can be found in the respective publications.

Two choices that are common across analyses and have direct impacts on displaced objects are the jet vertex tagger requirement and the lepton identification requirement. To minimize the contribution from jets arising from pile-up interactions, the jets used by the analyses
must satisfy a loose jet vertex tagger~(JVT) requirement~\cite{PERF-2014-03}. The JVT algorithm uses tracking and primary vertex information to determine if a given jet originates from the primary vertex.
The chosen working point has an efficiency of 94\% at a jet \pT\ of 40~\GeV\ and is
nearly fully efficient at 60~\GeV\ for jets originating from the primary vertex.
The JVT requirement is only applied to jets up to 60~\GeV\ and within $|\eta|<2.4$.\footnote{The RPV 1L analysis applies JVT to all jets within $|\eta|<2.4$, regardless of their \pt. }
Jets above this \pt threshold will be accepted by the analyses even if they originate from displaced decays.

The leptons used by the different analyses have requirements on their transverse impact parameter ($d_0$) and longitudinal impact parameter ($z_0 \sin \theta$) in the final state.
The muon (electron) identification algorithms require $|z_0 \sin \theta| < 0.5$ mm and $|d_0|/\sigma_{d_0} < 3~(5)$ except for the DL analysis, where these requirements were optimized for displaced leptons.
Therefore, only the DL analysis selects significantly displaced leptons, for which dedicated uncertainties are described in~\cite{SUSY-2018-14}.

The performance of the reconstruction and calibration algorithms on displaced signals is studied, and dedicated uncertainties are developed to account for possible mismodelling in the MC simulation of such topologies.
For each analysis, these uncertainties are considered in addition to the full set of uncertainties described in their respective publications.
These uncertainties are only applied to signal samples, as the background events in all signal regions originate from prompt processes. Furthermore, correlations of systematic uncertainties across different analyses are not considered, as each search is reinterpreted independently.

\subsection{Jet energy scale uncertainties for displaced jets}
Given the lack of a suitable data sample containing displaced jets, an MC-based prescription is designed to evaluate additional uncertainties for displaced jets.
The jet response, defined as the ratio of reconstructed jet~\pt over the truth jet~\pt, is studied in order to understand the effects of jet displacement on the jet energy scale (JES) and the jet energy resolution (JER).

The procedure of constructing and investigating the jet response follows the strategy described in Ref.~\cite{PERF-2016-04}.
The jet response is computed from reconstructed jets geometrically matched to truth jets using the distance measurement ($\Delta R$) .
Jets referred to as \textit{truth} jets are reconstructed from stable final-state generator-level particles in the simulation samples, using the same anti-$k_t$ algorithm~\cite{Cacciari:2008gp,Fastjet} with $R=0.4$ as used for reconstructed jets.
Muons, neutrinos, and any particles originating from pile-up interactions are excluded from the input used to build truth jets.
The simulation samples contain decaying neutralinos with lifetimes ranging from 0.01 ns to 100 ns.
Truth jets are labeled as originating from a long-lived particle by performing a \pt-dependent $\Delta R$ matching to the decaying neutralino. Since reconstructed jets are always assumed to originate from the primary vertex, a geometric correction to $\theta$ and $\phi$ is applied to the direction of the reconstructed jets, according to the position of the displaced vertex from the long-lived particle. This correction is performed only to improve the matching to truth jets, and the jet response is computed with respect to the uncorrected jet.
Only isolated jets are used to compute the jet response, not only to mitigate the influence of near-by jets but also to ensure consistency with the object definitions used in the analysis. Reconstructed jets are required to have no additional reconstructed jets within a cone of $\Delta R=0.6$, and exactly one truth jet within a cone of $\Delta R=1.0$.

The deviation from unity of the observed jet response is affected by the volume of the calorimeter traversed. %
A conservative approach is adopted in which the jet response uncertainty is assigned as the full difference in the response as a function of the radial decay length of the long-lived particle. The assigned uncertainty is below the percent level for radial decay lengths below 1~m, corresponding to decays occurring before reaching the calorimeter, then grows linearly reaching 30\% at 1.6~m, and remains approximately constant until the radial decay reaches the outside surface. Several studies of jets and calorimeter clusters in data~\cite{PERF-2016-04, EXOT-2017-25} with properties similar to displaced jets have shown much smaller levels of disagreement than the uncertainties assigned in this analysis. The jet-level uncertainties are translated to event-level uncertainties based on the expected jet response for the transverse displacements of the neutralinos decays.
The RPV multijet analysis is the most sensitive to such effect.
For Gqq models, the resulting uncertainty on event yields is close to 5\% at an LSP lifetime of $0.1\,$ns and increases up to 50\%  when the lifetime is $1\,$ns.
The event yields changes by less then $2\%$ for LSP lifetimes of $0.1\,$ns and below,
while it drops by almost $35\%$ at $1\,$ns.
For even more displaced decays, this uncertainty reduces again since the \ninoone escapes the calorimeters in most of the events.

Similarly to what is the done for JES, an extra systematic for JER is considered by studying the evolution of the width of the jet response as a function of the radial decay length.
However, the variation of the JER is smaller than the uncertainty associated to it, and is not considered as an additional systematic.

\subsection{\textit{b}-tagging uncertainties for displaced jets}

The $b$-tagging efficiency for displaced jets differs from that of prompt jets originating at the primary vertex.
For displaced $b$-jets, the efficiency depends on the alignment between the $B$-hadron momentum and the long-lived particle flight direction.
This alignment is quantified by the collinearity variable
\begin{equation}
IP_B = L'_{xy}\sin(\Delta\phi'),
\end{equation}
where $L'_{xy}$ is the transverse displacement of the $B$-hadron decay vertex and $\Delta\phi'$ is the azimuthal angle between the $B$-hadron momentum and its displacement vector from the primary vertex.
An analogous quantity, $IP_q$, is defined for light-flavor jets using the LLP decay position ($L_{xy}$) and the light-quark momentum. Jets with a small collinearity variable are classified as \textit{in-core}, while jets with a large collinearity variable are classified as \textit{off-core}\footnote{Displaced $b$-jets are classified as in-core if $|IP_B|<0.5~\mathrm{mm}~(L_{xy}<20~\mathrm{mm})$, or $|IP_B|<10~\mathrm{mm}^2/L_{xy}~(L_{xy}\geq20~\mathrm{mm})$, and are classified as off-core otherwise. Displaced light-flavor jets are classified as in-core if $|IP_q|<5$~mm, and are classified as off-core otherwise}.

\begin{figure}[!ht]
\centering
\subfloat[]{\includegraphics[width=0.52\textwidth]{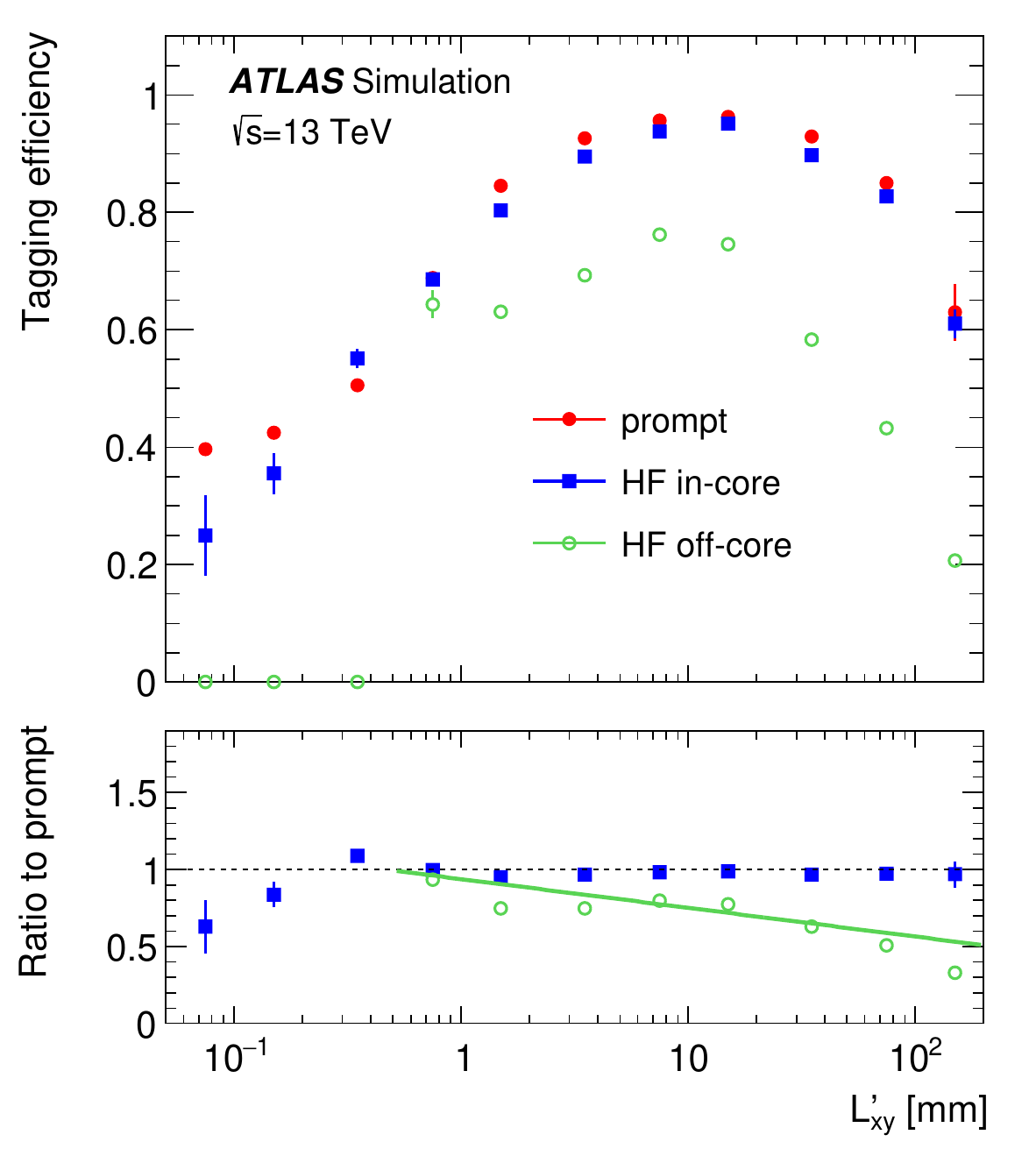} \label{fig:bjet_frac}}
\subfloat[]{\includegraphics[width=0.52\textwidth]{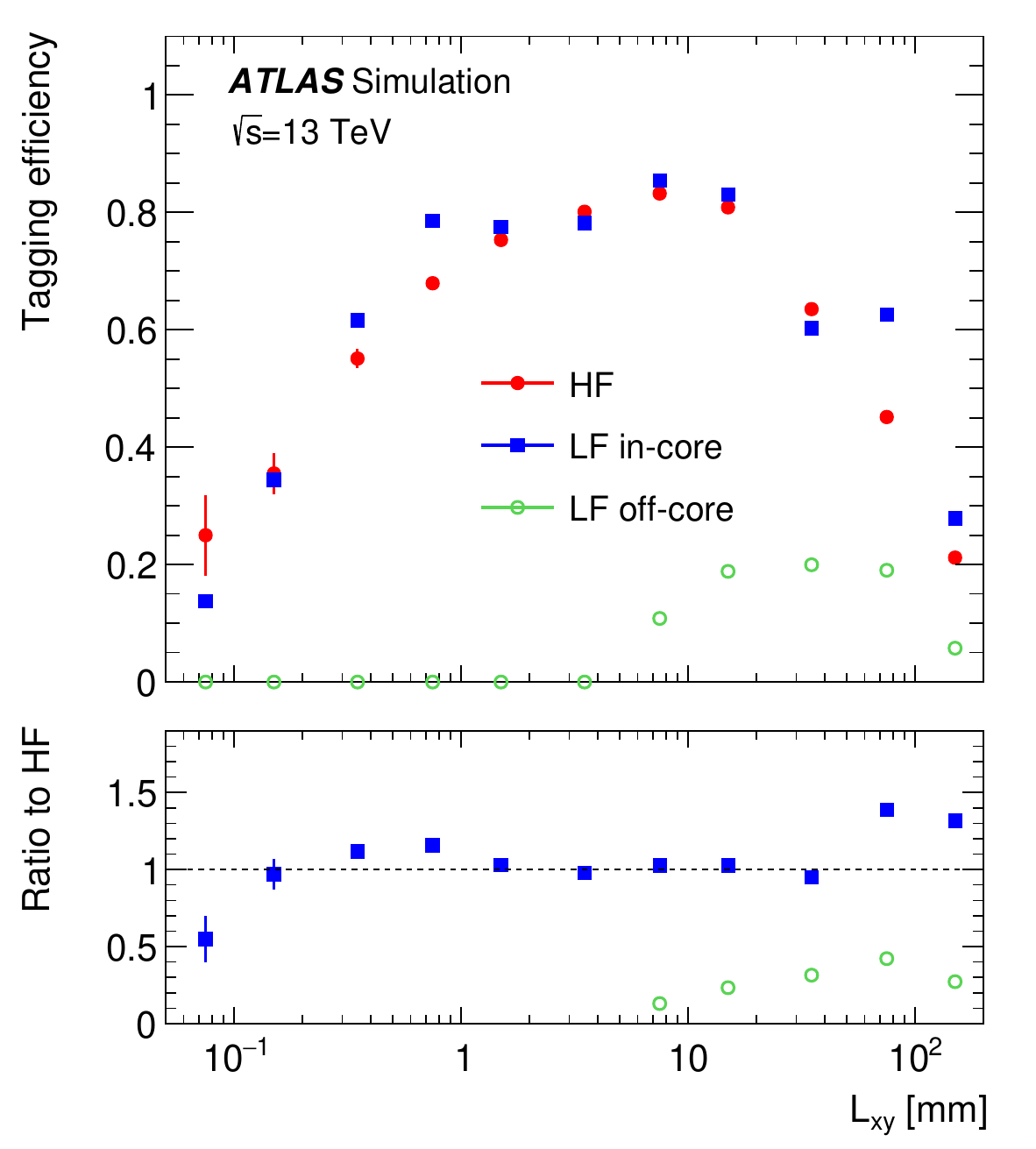} \label{fig:ljet_frac}}
\caption{The $b$-tagging efficiency for \protect\subref{fig:bjet_frac} displaced $b$-jets (HF) as a function of the position of the $B$-hadron decay vertex in the transverse plane ($L^{\prime}_{xy}$) and \protect\subref{fig:ljet_frac} displaced light-flavor jets (LF) as a function of the transverse LLP decay position ($L_{xy}$). The efficiency is shown separately for jets where the $B$-hadron or light-flavor quark direction is collinear (\textit{in-core}) or is not collinear (\textit{off-core}) with the LLP direction. The prompt sample shown in \protect\subref{fig:bjet_frac} is from $b$-jets in a $t\bar{t}$ sample, while the HF shown in \protect\subref{fig:ljet_frac} is a combination of in-core and off-core displaced HF jets.
The bottom panel shows the ratio of the $b$-tagging efficiencies for  \protect\subref{fig:bjet_frac} displaced HF jets and prompt $b$-jets in a $t\bar{t}$ sample, and for \protect\subref{fig:ljet_frac} displaced light-flavor jets to displaced HF jets. The function drawn in \protect\subref{fig:bjet_frac} is used to derive the uncertainties for off-core HF jets.
}
\end{figure}

The $b$-tagging efficiency of displaced in-core $b$-jets, shown in Figure~\ref{fig:bjet_frac}, is comparable to that of prompt $b$-jets with similar decay lengths.
As the $IP_B$ increases, the efficiency decreases due to the loss of displaced tracks passing the impact-parameter requirements of the flavor-tagging algorithm.
The $b$-tagging efficiency for displaced in-core light-flavor jets, shown in Figure~\ref{fig:ljet_frac}, resembles that of $b$-jets, because the tracks inside the light-flavor jet acquire non-zero impact parameters similar to tracks from $B$-hadron decays.
At larger $IP_q$, the tagging efficiency decreases, similar to that of displaced $b$-jets.

The \textit{b}-tagging uncertainties for displaced jets are evaluated for the ATLAS flavor tagging algorithm \texttt{MV2c10}~\cite{FTAG-2018-01} and working point \texttt{WP77}, as used by the multi-$b$ analyses.
An additional per-jet uncertainty ($\sigma_{\mathrm{add}}$) is added in quadrature with the standard $b$-tagging uncertainties.
For displaced $b$-jets, the uncertainty is derived by comparing the $b$-tagging efficiency for displaced $b$-jets to prompt $b$-jets in a sample of simulated $t\bar{t}$ events events, and is set to $\sigma_{\mathrm{add}} = 2.2\%$ for in-core jets and $\sigma_{\mathrm{add}} = 8\ln(L_{xy}[\mathrm{mm}]/0.5)\%$ for off-core jets.
For displaced light-flavor jets, the uncertainties are derived by comparing the $b$-tagging efficiency for displaced light-flavor jets and displaced $b$-jets.
The uncertainty $\sigma_{\mathrm{add}}$ is taken to be 1.6\% for in-core jets and conservatively set to 100\% of the $b$-tagging efficiency for off-core jets.

The RPV multi-$b$ analysis is the most sensitive to $b$-tagging uncertainties due to the many $b$-jets in its final states. For Gtt models, the standard $b$-tagging uncertainties of light-jets \cite{FTAG-2018-01} originating from LLP decays propagated to an event yield uncertainty reaches as high as 40\% for an LSP lifetime of $0.01\,$ns. This uncertainty is reduced to 9\% when the dedicated $b$-tagging uncertainty estimation is applied. For longer lifetimes, the dedicated uncertainty is about 5\% at an LSP lifetime of $100\,$ns. Despite these changes, the effect on the exclusion limits is minimal, corresponding to an extension of about 50 GeV compared to the standard approach.

\subsection{\texorpdfstring{$\tau$}{τ}-lepton uncertainties for displaced taus}
The efficiency of $\tauhad$ reconstruction and identification is expected to be affected by the additional decay length induced by the long-lived particle.
The reconstruction and identification efficiency for $\tauhad$ candidates decreases rapidly when the impact parameter $d_0$ of the $\tauhad$ is larger than a few millimeters, while the dependence on the \tauhad \pT and $|\eta|$ is comparatively weaker.
Therefore, the efficiency is parameterised as a function of the \tauhad $d_0$ and $L_{xy}$ separately for one- and three-prong $\tauhad$.
The ranges from $10^{-5}$~mm to $10$~mm in $d_0$ and $10^{-2}$~mm to $10^{3}$~mm in $L_{xy}$ are studied.
The systematic uncertainties are taken to be the same as the ones documented in Ref.~\cite{ATL-PHYS-PUB-2024-007}. %
The efficiency differences between $\tauhad$ candidates in $Z\to\tau\tau$ events and simulated RPV signal events are assigned as systematic uncertainties.
The uncertainties are evaluated for the \texttt{Medium} $\tauhad$ working point, which corresponds to a signal efficiency of 75\% (60\%) for 1-prong (3-prong) $\tauhad$~\cite{ATL-PHYS-PUB-2019-033}.
The resulting uncertainties are estimated to be between 1\% and 5\% for 1-prong $\tau$ and 1\% to 20\% for 3-prong $\tau$ depending on the values of $d_0$ and $L_{xy}$.

The trigger efficiency for the di-$\tau$ and di-$\tau$+$\MET$ triggers used in the di-$\tau$ analysis is parameterized as a function of the sub-leading \tauhad \pT.
The systematic uncertainty is defined as the difference in efficiency between $Z\to\tau\tau$ events and the simulated RPV signal events.
The di-$\tau$ trigger efficiency is flat as a function of sub-leading \tauhad \pT and the corresponding efficiency difference is within 6\%.
For the di-$\tau$+$\MET$ trigger, the efficiency flattens for $\MET$ above 130 GeV and the corresponding difference is within 12\%.

After including all relevant additional $\tauhad$ uncertainties, the Higgsino model is the one most impacted by the uncertainties. The exclusion limit obtained by the RPC Di-$\tau$ analysis reduces by up to 0.02 ns in the LSP lifetime.

\subsection{Missing transverse momentum uncertainties}
The missing transverse momentum is reconstructed as the negative vector sum of the transverse momenta of the hard objects in the event, together with a soft term built from high-quality tracks associated with the
primary vertex but not with any physics objects~\cite{JETM-2020-03}. Variations on the hard objects due to systematic uncertainties are propagated to the \met\ . The soft term uses only tracks associated with the primary vertex and therefore does not introduce additional uncertainties for displaced signals. Studies in simulation show no significant dependence of the \met\ trigger efficiency turn-on on the LSP lifetime, and no additional trigger systematic uncertainty is assigned.


%
\section{Results}
\label{sec:result}

Results are provided in the context of six RPV SUSY benchmark models discussed in Section~\ref{sec:susymodels},
using the thirteen ATLAS analyses described in Section~\ref{sec:analyses}.
In all cases except for the dijet analyses, the profile likelihood-ratio test~\cite{Cowan:2010js} is used to establish 95\% confidence intervals using the CL$_\text{s}$ prescription. In the dijet analyses, a Bayesian procedure is used to set 95\% confidence-level upper limits on generic Gaussian resonances~\cite{EXOT-2019-03}.
Individual limits from each analysis are reported, and with the exception of the RPV stop dijet and RPV dijet resonance analyses, no combination is performed due to substantial overlaps in signal region definitions. %

\begin{figure}[!ht]
\centering
\includegraphics[width=0.6\textwidth]{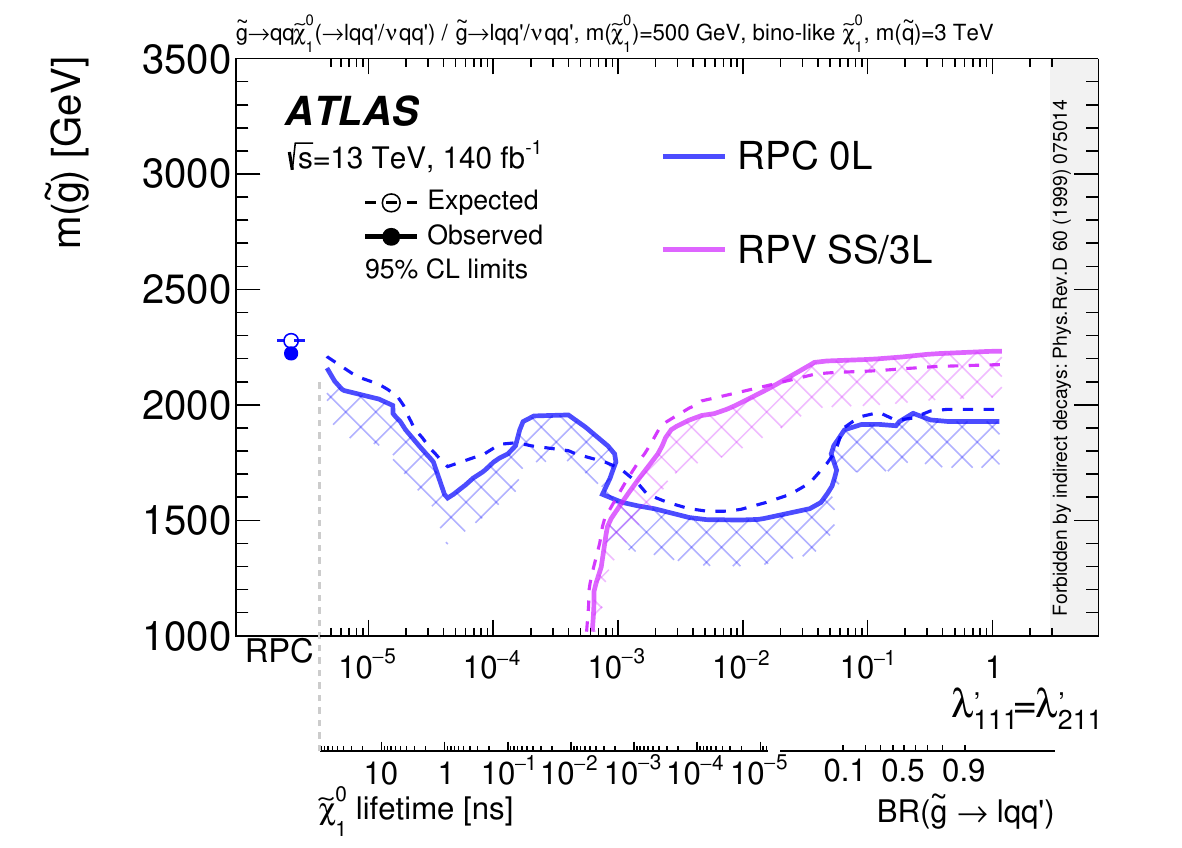}
\caption{Exclusion limits for the Gqq+LQD model as a function of $\lambdap_{111}$ and $m(\gluino)$ under the assumption that $\lambdap_{111}=\lambdap_{211}$. Expected limits are shown with dashed lines, and observed as solid. The RPC-limit is shown on the leftmost part of the axes. The region $\lambda^{\prime}_{111}$ greater than 2.91 is forbidden by constraints on the $0\nu\beta\beta$ measurements~\cite{Allanach:1999ic}. The exclusion limits are derived assuming the squark masses listed in Table~\ref{tab:Modelsummary}.}
\label{fig:gqqlqd}
\end{figure}

Figure~\ref{fig:gqqlqd} shows the observed and expected gluino mass lower limits obtained in the Gqq+LQD model, as a function of the neutralino lifetime and the equivalent $\lambdap_{111}$ coupling strength, under the assumption that $\lambdap_{111}=\lambdap_{211}$. The RPC 0L analysis sets the strongest limits in the low $\lambdap_{111}$ regime, where the LSPs are sufficiently long-lived to contribute to \met. The RPV SS/3L analysis on the other hand sets strong limits in the large $\lambdap_{111}$ regime. Gluino masses up to 2.2~TeV are excluded for a neutralino lifetime around 100 ns. The fluctuating behaviour of the RPC 0L curve is due to variations in the fraction of leptons that fail reconstruction and consequently contribute to \met, which modulates the sensitivity across lifetimes. The dip at 1 ns arises from the lower \met observed at this lifetime compared with 0.1 ns and 0.01 ns. The kink near 0.01 ns originates from the combination of limits from different signal regions when using the expected CLs values.

\begin{figure}[hptb]
\centering
\includegraphics[width=0.7\textwidth]{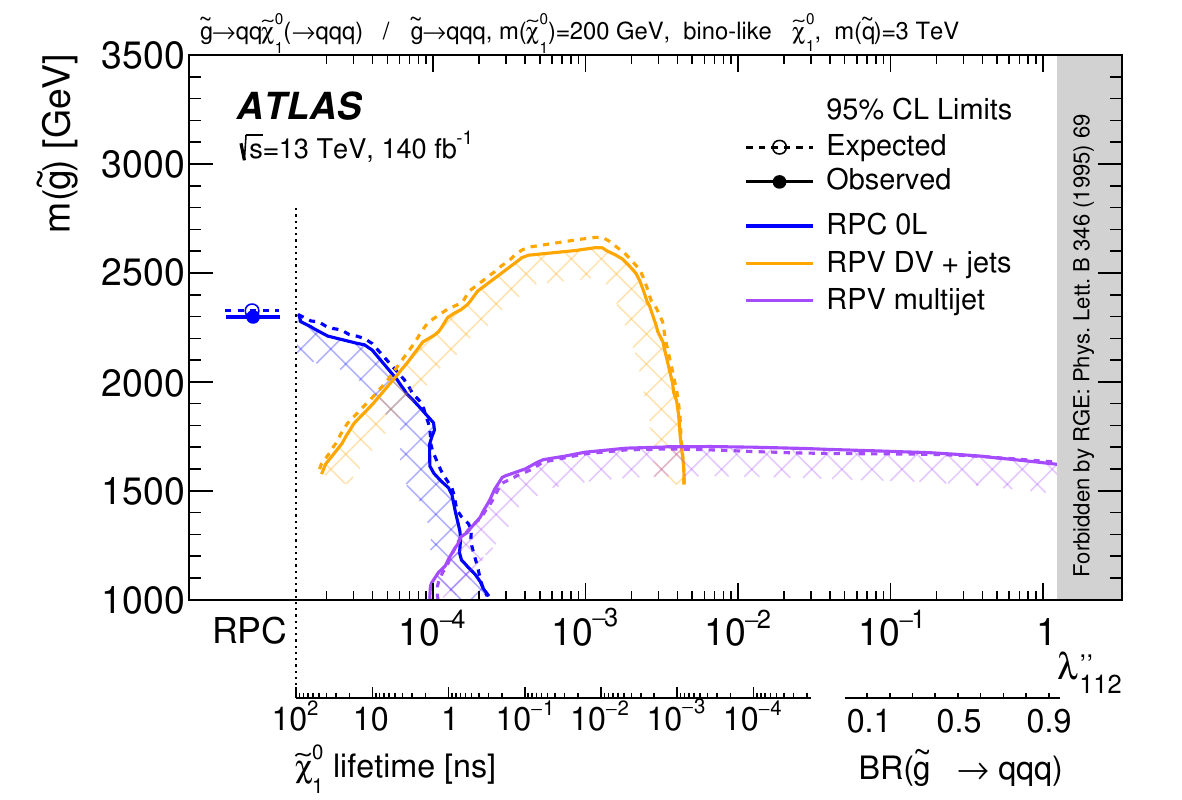}
\caption{Exclusion limits for the Gqq+UDD model as a function of $\lambdapp_{112}$ and $m(\gluino)$. Expected limits are shown with dashed lines, and observed as solid.
The RPC-limit is shown on the leftmost part of the axes. The region $\lambdapp_{112}>1.25$ is forbidden by constraints from the renormalization group equations. The exclusion limits are derived assuming the squark masses listed in Table~\ref{tab:Modelsummary}.} %
\label{fig:all_gqq}
\end{figure}

Figure~\ref{fig:all_gqq} shows the observed and expected gluino mass lower limits obtained in the Gqq+UDD model,
as a function of the neutralino lifetime and the equivalent $\lambdapp_{112}$ coupling strength. The DV+jets analysis, which was optimized for this model, sets the strongest mass limit at $\lambdapp_{112}\approx10^{-3}$, corresponding to $\tau_\text{LSP}\approx$~0.1 to 0.01~ns. The sensitivity of the DV+jets analysis is limited at higher and lower values of the $\lambdapp_{112}$ coupling, when the  \ninoone decays occur promptly or beyond the volume of the inner detector, respectively. The RPC 0L analysis also sets strong limits on this model in the small $\lambdapp_{112}$ regime as expected: these signals generate the substantial \met which the analysis uses to discriminate between signal and background.
The sensitivity of the analysis falls off rapidly as the lifetime of the \ninoone decreases and the decays to jets reduce the \met. The RPV Multijet analysis has the highest sensitivity at medium to large $\lambdapp_{112}$ regime with LSP or direct RPV gluino decays. Gluino masses up to 2.6~\tev\ are excluded for a neutralino lifetime of 0.01~ns, and up to 2.3~\tev\ for a lifetime of 100 ns. Previous searches in all-hadronic final states have set a limit of about 2.2~\tev\ on the gluino mass when considering only the prompt decay $\gluino \rightarrow qq \ninoone$~\cite{SUSY-2019-24}. While the generated model has non-zero values only for $\lambdapp_{112}$, the limits apply for any $\lambdapp_{ijk}$ with $i \neq 3$.

\begin{figure}[hptb]
\centering
\includegraphics[width=0.7\textwidth]{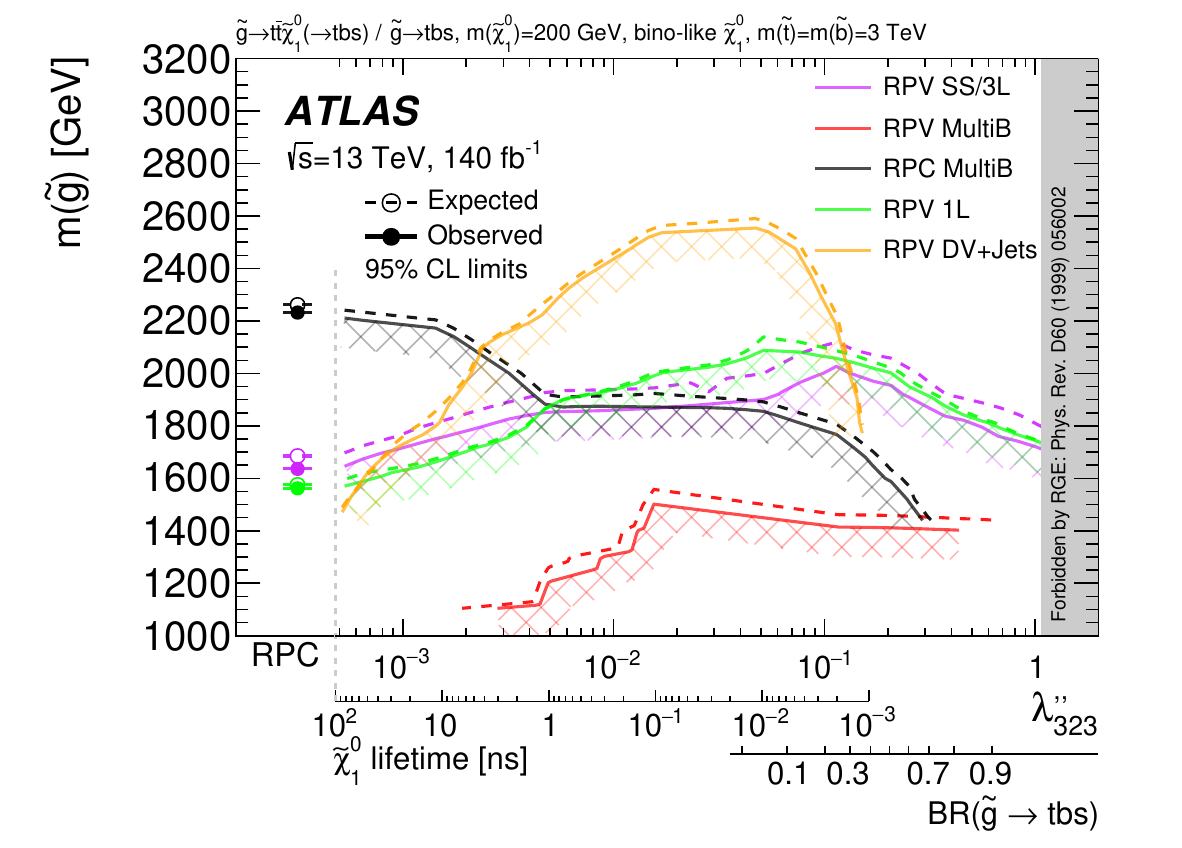}
\caption{Exclusion limits for the Gtt model as a function of $\lambdapp_{323}$ and $m(\gluino)$. Expected limits are shown with dashed lines, and observed as solid.
The region $\lambdapp_{323} > 1.07$ is forbidden by constraints from the renormalization group equations. The exclusion limits are derived assuming the squark masses listed in Table~\ref{tab:Modelsummary}.
}
\label{fig:all_gtt}
\end{figure}

Figure~\ref{fig:all_gtt} shows the observed and expected gluino mass lower limits obtained in the Gtt model,
as a function of the neutralino lifetime and the gluino branching fraction, as well as the equivalent $\lambdapp_{323}$ coupling strength.
The entire $\lambdapp_{323}$ range is covered effectively by these analyses.
The sensitivity of the analyses optimised for RPV signals increases with larger $\lambdapp_{323}$ as more \ninoone decays occur inside the detectors due to the decreasing $\tauLSP$, leading to more reconstructed jets in the final states.
However, the limits start degrading when the \gluino decay transitions from the \ninoone cascade to direct RPV decay, reducing the jet and top-quark multiplicities. %
For $\lambdapp_{323}\approx 1$ the \gluino decays directly to fewer jets, leading to worse signal-background separation and a reduced sensitivity. The RPC multi-$b$ analysis sets the strongest limit in RPC SUSY scenarios and at lower values of $\lambdapp_{323}$ because the LSPs are sufficiently long-lived to contribute to the \met\ in the final state. The DV+jets analysis sets the strongest limits for moderate values of $\lambdapp_{323}$ with peak sensitivity achieved for signals with $\tauLSP \approx 0.1$ to $0.01$ ns.
The RPV multi-$b$ analysis sets the weakest limits because it relies on a cut-and-count strategy based solely on the jet and $b$-tagged jet multiplicities. In addition, its sensitivity is limited by the dominant QCD multijet background.

\begin{figure}[hptb]
\centering
\includegraphics[width=0.7\textwidth]{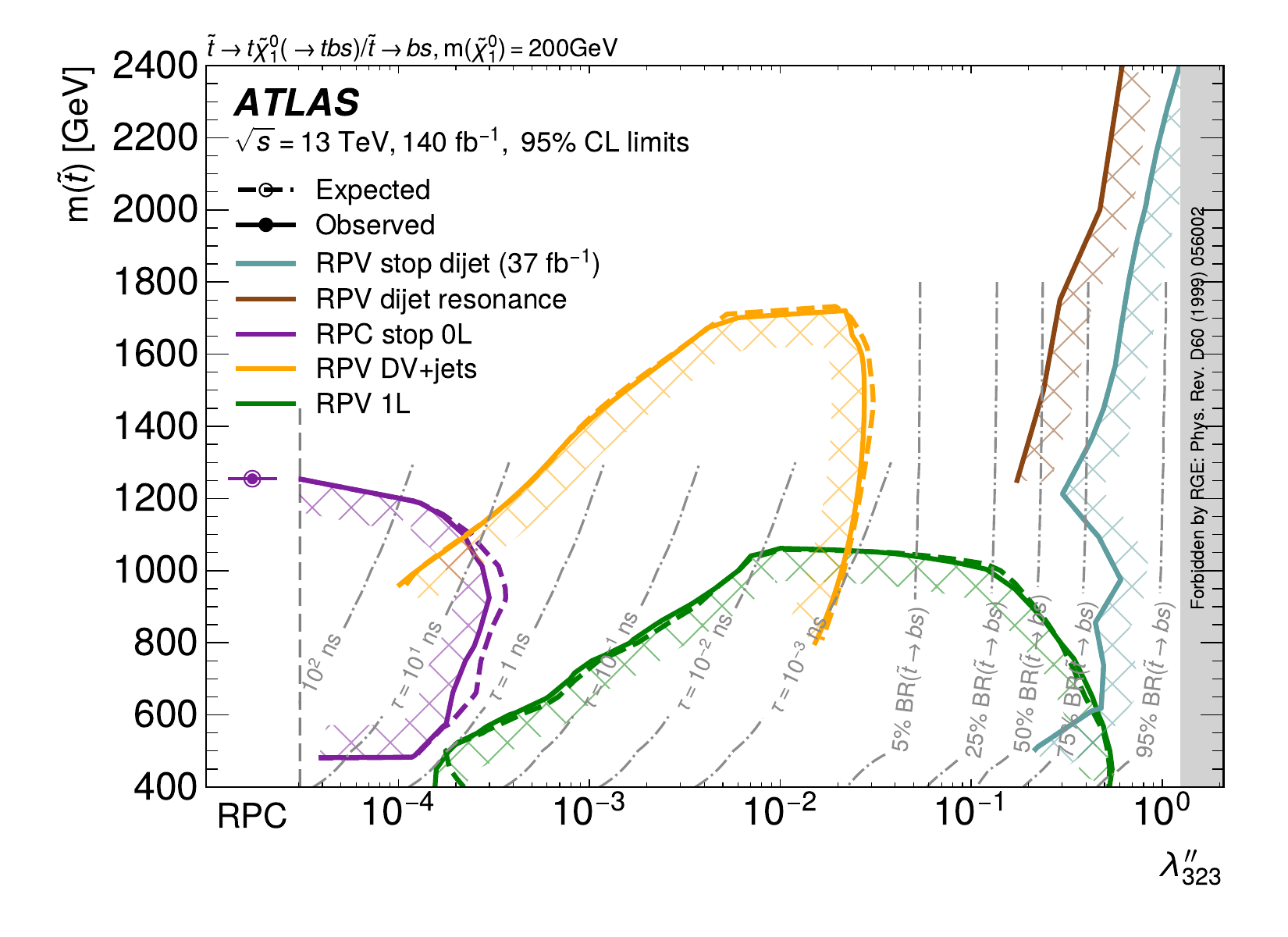}
\caption{Exclusion limits for the stop model as a function of $\lambdapp_{323}$ and $m(\stop)$. Expected limits are shown with dashed lines, and observed as solid.
The region $\lambdapp_{323} > 1.07$ is forbidden by constraints from the renormalization group equations. The exclusion limits are derived assuming the squark masses listed in Table~\ref{tab:Modelsummary}.}
\label{fig:all_stop}
\end{figure}

Figure~\ref{fig:all_stop} shows the observed and expected lower top-squark mass limits obtained in the stop model
as a function of $\lambdapp_{323}$.
Contours of $\ninoone$ lifetime and branching fractions of direct decays, $\stop \rightarrow bs$, are overlaid since they depend on both the coupling value and the top-squark mass.
In the RPC regime, and for low values of $\lambdapp_{323}$, the RPC stop 0L analysis sets the strongest limits.
The reliance on high \met signatures quickly reduces the sensitivity at even moderate values of $\lambdapp_{323}$ and \ninoone lifetime.
The RPV 1L analysis sets limits starting for slightly higher values of $\lambdapp_{323}$ and provides unique exclusion limits for $\lambdapp_{323}\approx 0.02 - 0.1$.
The RPV DV+jets analysis covers the gap between the RPC stop 0L and RPV 1L analyses and sets the strongest limits around $\lambdapp_{323}\approx 2\times10^{-4}-2\times10^{-2}$.
At very high values of $\lambdapp_{323}$, the RPV stop dijet and the RPV dijet resonance analyses set the strongest limits, which exclude the top-squark mass beyond 2.4 TeV at $\lambdapp_{323} \approx 1$.

\begin{figure}[hptb]
\centering
\subfloat[]{\includegraphics[width=0.495\textwidth]{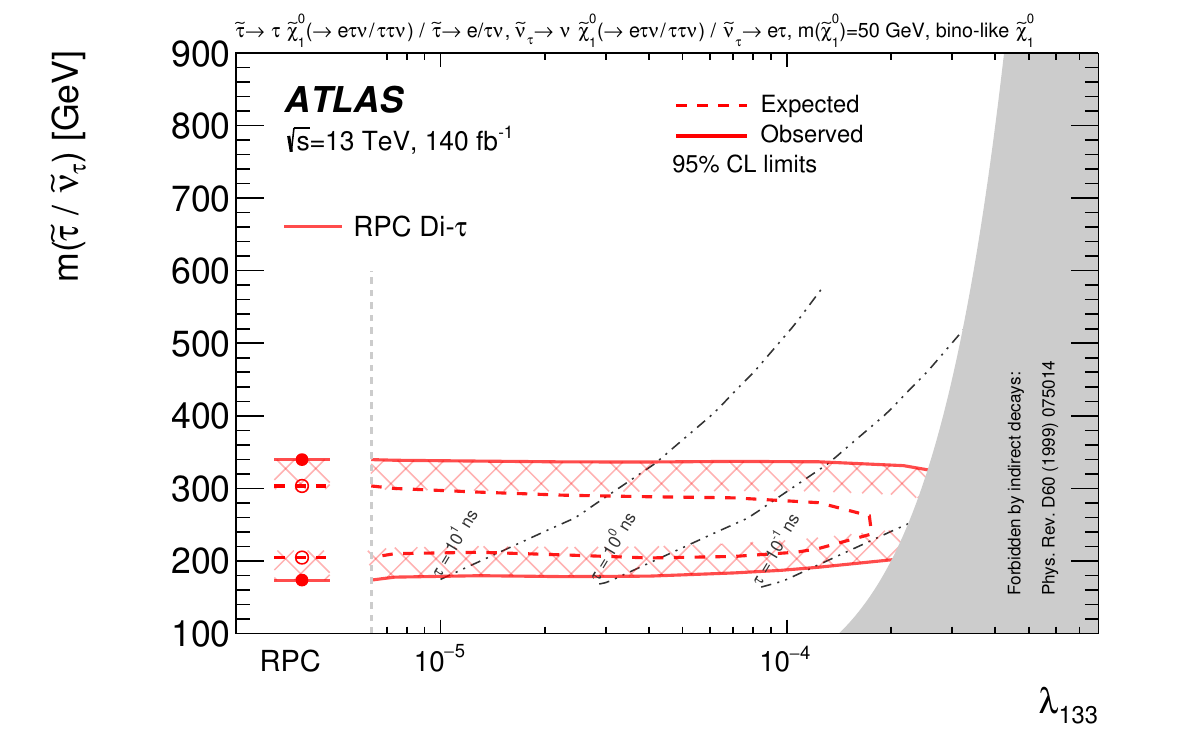}\label{fig:stau-133}}
\subfloat[]{\includegraphics[width=0.495\textwidth]{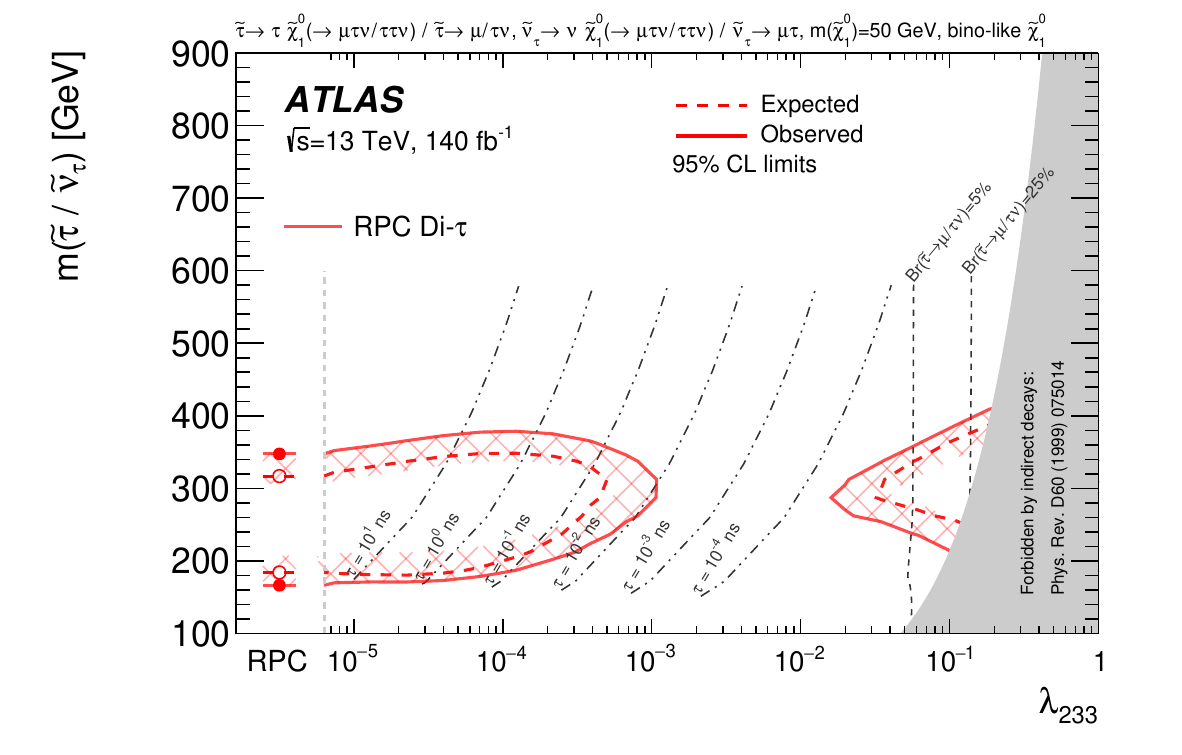}\label{fig:stau-233}}\\
\caption{Exclusion limits for the stau model as a function of $\tau$-slepton mass and $\lambda_{i33}$, for \protect\subref{fig:stau-133} $\lambda_{133}$ and \protect\subref{fig:stau-233} $\lambda_{233}$. Expected limits are shown with dashed lines, and observed as solid. The gray regions represents forbidden values by constraints on \protect\subref{fig:stau-133} m($\nu_{e}$)~\cite{KATRIN:2021uub} and \protect\subref{fig:stau-233} $R_{\tau}$ measurements~\cite{Workman:2022ynf}. The exclusion limits are derived assuming the slepton masses listed in Table~\ref{tab:Modelsummary}.}
\label{fig:stau}
\end{figure}

Figure~\ref{fig:stau} shows the observed and expected limits on the $\tau$-slepton mass for the stau model, as a function of $\lambda_{133}$ and $\lambda_{233}$ separately.
The sensitivity of the Di-$\tau$, 4L, and DL analyses to the stau model were evaluated; however, only the Di-$\tau$ analysis has sensitivity to exclude parts of the parameter space.
$\tau$-slepton masses between 180 \GeV\ and 340 \GeV\ are excluded for $\lambda_{i33} \leq 10^{-3}$, where the \ninoone decay products fail to be reconstructed, resulting in a RPC-like signature similar to that targeted by the original Di-$\tau$ analysis.
Exclusion limits are also set at very large values of $\lambda_{233}$, shown in Figure~\ref{fig:stau-233}, where the $\stau$ RPV direct decay dominates. In this region, $\stau$ can promptly decay to neutrinos and $\tau$-leptons, the presence of $\tau$-leptons and large \MET\ from neutrinos mimic the RPC di-$\stau$ final state.
In these excluded regions, the best sensitivity comes from the signal region requiring exactly two \tauhad with opposite sign charge of the Di-$\tau$ analysis.
In the intermediate $\lambda_{233}$ regime ($ 10^{-3} \leq \lambda_{233} \leq 10^{-1}$) of Figure~\ref{fig:stau-233}, where \ninoone decays are prompt enough to lead to a high $\tau$-lepton multiplicity, events fail the large \MET selection and the opposite-sign-charge requirement. This results in an overall loss of sensitivity for the Di-$\tau$ analysis.
The excluded region for $\lambda_{133}$ is smaller than that for $\lambda_{233}$.
This is because displaced electrons can be mis-reconstructed as photons or jets, thereby reducing the reconstructed \MET.
This leads to a lower \MET selection efficiency in electron final states compared to those involving muons.
The 4L analysis has no sensitivity due to the higher $\meff$ requirement, and the DL analysis lacks sensitivity due to the relatively high \pT requirement for displaced leptons.

\begin{figure}[hptb]
\centering
\subfloat[]{\includegraphics[width=0.495\textwidth]{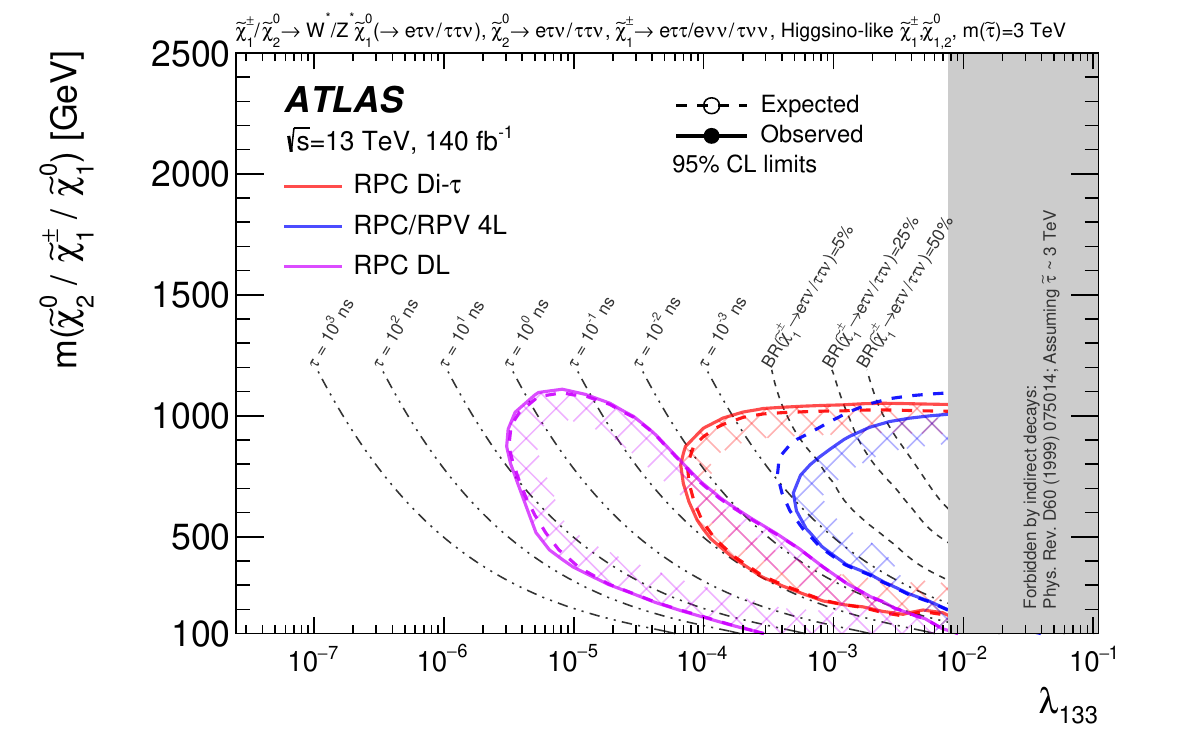}\label{fig:higgsino-133}}
\subfloat[]{\includegraphics[width=0.495\textwidth]{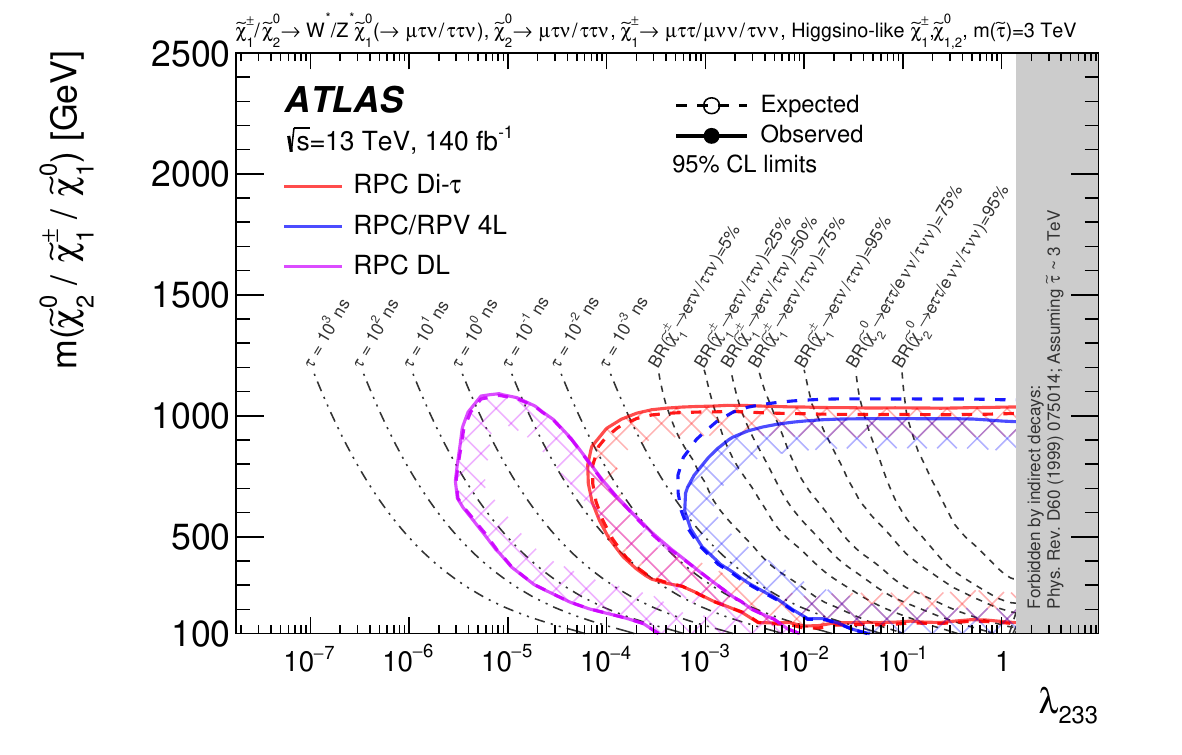}\label{fig:higgsino-233}}\\
\caption{ Exclusion limits for the higgsino model as a function of higgsino mass and \protect\subref{fig:higgsino-133} $\lambda_{133}$, or \protect\subref{fig:higgsino-233} $\lambda_{233}$ for the combined production of $\tilde{\chi}_1^0 \tilde{\chi}_2^0$, $\tilde{\chi}_1^0 \tilde{\chi}_1^{\pm}$ and $\tilde{\chi}_1^{\pm} \tilde{\chi}_1^{\mp}$. Expected (observed) limits are shown with dashed (solid) lines.
The region $\lambda_{133}$ > 0.0077 ( $\lambda_{233}$ > 1.42 ) is forbidden by constraints on m($\nu_{e}$)~\cite{KATRIN:2021uub} measurements ($R_{\tau}$ measurements~\cite{Workman:2022ynf}).
The exclusion limits are derived assuming the slepton masses listed in Table~\ref{tab:Modelsummary}.
}
\label{fig:higgsino}
\end{figure}

Figure~\ref{fig:higgsino} shows the observed and expected limits on the higgsino mass for the production of $\tilde{\chi}_1^0 \tilde{\chi}_2^0$, $\tilde{\chi}_1^0 \tilde{\chi}_1^{\pm}$, and $\tilde{\chi}_1^{\pm} \tilde{\chi}_1^{\mp}$ as a function of $\lambda_{133}$ and $\lambda_{233}$ separately.
The largest sensitivity for both Di-$\tau$ and the 4L analyses occurs at large $\lambda_{133}$ and $\lambda_{233}$ values, where the $\tilde{\chi}_1^0$, $\tilde{\chi}_2^0$ and $\tilde{\chi}_1^{\pm}$ decay promptly into light leptons, $\tau$-leptons, and neutrinos.
In these regions, the higgsino mass exclusion reach up to 1 \TeV. The sensitivity of the 4L analysis decreases rapidly as $\tau_\text{LSP}$ approaches 10$^{-4}$ ns, since the electrons and muons from the $\tilde{\chi}_1^0$ decay start to fail the prompt-lepton reconstruction requirements.
The most sensitive signal region in the 4L analysis for $\tilde{\chi}_1^0 \tilde{\chi}_2^0$ ($\tilde{\chi}_1^{\pm} \tilde{\chi}_1^{\mp}$) production requires two (three) leptons and $m_{\mathrm{eff}} > 600\ \GeV$. For $\tilde{\chi}_1^0 \tilde{\chi}_1^{\pm}$ production, the signal region requiring two leptons and $m_{\mathrm{eff}} > 1000\ \GeV$ is the most sensitive one at high higgsino masses, while for higgsino masses around 200~\GeV, the most sensitive region requires at least four leptons, $\MET > 100\ \GeV$, and two \Zboson-boson candidates.
Given the large $\tau$-lepton multiplicity in the models, the most sensitive signal region for the Di-$\tau$ analysis requires at least two medium \tauhad and $\MET > 150$~\GeV.
The Di-$\tau$ analysis excludes higgsino masses from 200 \GeV\ to 1 \TeV, with sensitivity decreasing rapidly for $\lambda_{133}$ and $\lambda_{233}$ values corresponding to $\tau_\text{LSP}\sim 0.01$~ns, when the \tauhad reconstruction efficiency drops.
For the DL analysis, the higgsino masses up to 1.1 \TeV are excluded for $\lambda_{133}$ ($\lambda_{233}$) $\sim 5\times10^{-5}$. The most sensitive signal regions in the DL analysis require two same-flavor leptons. Overall, the exclusion limits on $\tau_\text{LSP}$ reach up to approximately $10$~ns for the higgsino combined production in both $\lambda_{133}$ and $\lambda_{233}$ scenarios.
Constraints on the higgsino mass as a function of $\Delta m(\tilde{\chi}_1^{\pm},\ninoone)$ have been established by other ATLAS searches using the soft lepton ~\cite{SUSY-2019-09,SUSY-2018-16}, displaced track~\cite{SUSY-2020-04,HMBS-2024-65}, and disappearing track~\cite{SUSY-2018-19}. These results are not incorporated into the current reinterpretation.
Furthermore, ATLAS searches for heavy neutral leptons featuring DV~\cite{EXOT-2022-12}, or searches for DV in the muon spectrometer~\cite{EXOT-2019-24,EXOT-2022-17} may provide sensitivity to longer $\tau_\text{LSP}$, but are not considered in this effort.

%

%
%
\FloatBarrier

\section{Conclusion}

This work provides a unified reinterpretation of published ATLAS SUSY searches across the full range of RPV coupling strengths, covering scenarios with prompt, long-lived, and effectively stable LSP decays. By incorporating the full Run-2 dataset, expanding the set of RPV SUSY models considered, and including long-lived particle searches for the first time in this context, this analysis significantly extends previous reinterpretation efforts and enables a more comprehensive assessment of the sensitivity of SUSY searches in ATLAS to signatures of RPV SUSY.

Six simplified models targeting RPV scenarios are analyzed.
The large variations in final state, and therefore in sensitivity, as a function of the $R$-parity coupling strength motivate a thorough examination of the ATLAS SUSY program's coverage from 13 dedicated searches.
Different degrees of sensitivity are observed as a function of  $\lambdap$, $\lambdapp$ or $\lambda_{i33}$ ($i \in \{1,2\}$): in the gluino model with large branching fractions to top quarks,
gluinos are excluded up to masses of 1.8 \TeV, over the full range of lifetime and RPV coupling strengths.
The gluino model with a $\lambdap$ coupling sets gluino mass limits above 1.6 \TeV, extenting the limit to 2.2 TeV for both high and low values of $\lambdap$. In contrast, the gluino model with decays to first- and second- generation quarks shows more pronounced variations: for low and medium values of the coupling $\lambdapp$, gluino masses are excluded up to 2.0--2.5 \TeV, whereas at high \lambdapp, gluino masses are only excluded up to 1.6 \TeV. The stop model shows large variations in the limits as well: top-squark masses up to 2.4 TeV can be excluded at high values of \lambdapp, whereas for lower \lambdapp the exclusion limits range from 1 TeV to 1.7 TeV, depending on the value of \lambdapp. The weakest exclusion limits are obtained for $\lambdapp \approx 0.1$.
$\tau$-slepton masses between 180 \GeV\ and 340 \GeV\ are excluded for $\lambda$ couplings less than~$10^{-4}$. The exclusion power drops considerably for~$ 10^{-3} \leq \lambda_{233} \leq 10^{-1}$. In higgsino scenarios with large coupling, where the higgsinos decay mainly via RPV couplings into SM particles, masses up to 800 GeV - 1 \TeV\ are excluded across all production channels.
For the higgsino model no exclusion limits can be established for values of $\lambda_{i33}$ smaller than~$3\times10^{-6}$.

\clearpage

\section*{Acknowledgements}

%
%

%
%

We thank CERN for the very successful operation of the LHC and its injectors, as well as the support staff at
CERN and at our institutions worldwide without whom ATLAS could not be operated efficiently.

The crucial computing support from all WLCG partners is acknowledged gratefully, in particular from CERN, the ATLAS Tier-1 facilities at TRIUMF/SFU (Canada), NDGF (Denmark, Norway, Sweden), CC-IN2P3 (France), KIT/GridKA (Germany), INFN-CNAF (Italy), NL-T1 (Netherlands), PIC (Spain), RAL (UK) and BNL (USA), the Tier-2 facilities worldwide and large non-WLCG resource providers. Major contributors of computing resources are listed in Ref.~\cite{ATL-SOFT-PUB-2026-001}.

We gratefully acknowledge the support of ANPCyT, Argentina; YerPhI, Armenia; ARC, Australia; BMWFW and FWF, Austria; ANAS, Azerbaijan; CNPq and FAPESP, Brazil; NSERC, NRC and CFI, Canada; CERN; ANID, Chile; CAS, MOST and NSFC, China; Minciencias, Colombia; MEYS CR, Czech Republic; DNRF and DNSRC, Denmark; IN2P3-CNRS and CEA-DRF/IRFU, France; SRNSFG, Georgia; BMFTR, HGF and MPG, Germany; GSRI, Greece; RGC and Hong Kong SAR, China; ICHEP and Academy of Sciences and Humanities, Israel; INFN, Italy; MEXT and JSPS, Japan; CNRST, Morocco; NWO, Netherlands; RCN, Norway; MNiSW, Poland; FCT, Portugal; MNE/IFA, Romania; MSTDI, Serbia; MSSR, Slovakia; ARIS and MVZI, Slovenia; DSI/NRF, South Africa; MICIU/AEI, Spain; SRC and Wallenberg Foundation, Sweden; SERI, SNSF and Cantons of Bern and Geneva, Switzerland; NSTC, Taipei; TENMAK, T\"urkiye; STFC/UKRI, United Kingdom; DOE and NSF, United States of America.

Individual groups and members have received support from BCKDF, CANARIE, CRC and DRAC, Canada; CERN-CZ, FORTE and PRIMUS, Czech Republic; COST, ERC, ERDF, Horizon 2020 and Marie Sk{\l}odowska-Curie Actions, European Union; Investissements d'Avenir Labex, Investissements d'Avenir Idex and ANR, France; DFG and AvH Foundation, Germany; Herakleitos, Thales and Aristeia programmes co-financed by EU-ESF and the Greek NSRF, Greece; BSF-NSF and MINERVA, Israel; NCN and NAWA, Poland; La Caixa Banking Foundation, CERCA and AGAUR programs from Generalitat de Catalunya and PROMETEO and GenT Programmes Generalitat Valenciana, Spain; G\"{o}ran Gustafssons Stiftelse, Sweden; The Royal Society and Leverhulme Trust, United Kingdom; Eric and Wendy Schmidt Fund for Strategic Innovation, United States of America.

In addition, individual members wish to acknowledge support from Chile: Agencia Nacional de Investigaci\'on y Desarrollo (ANID FONDECYT reg. 1230987, FONDECYT 1230812, FONDECYT 1240864, Fondecyt 3240661, Fondecyt Regular 1240721); China: Chinese Ministry of Science and Technology (MOST-2023YFA1605700, MOST-2023YFA1609300), National Natural Science Foundation of China (NSFC 12275265, NSFC-W2543005); Czech Republic: Czech Science Foundation (GACR - 24-11373S), Ministry of Education Youth and Sports (ERC-CZ-LL2327, FORTE CZ.02.01.01/00/22\_008/0004632), PRIMUS Research Programme (PRIMUS/21/SCI/017); EU: H2020 European Research Council (ERC - 101002463); European Union: European Research Council (BARD No. 101116429, ERC - 948254, ERC 101089007), European Regional Development Fund (HE COFUND GA No.101081355, ERDF), Marie Sklodowska-Curie Actions (GAP-101168829); France: Agence Nationale de la Recherche (ANR-21-CE31-0013, ANR-22-EDIR-0002, ANR-24-CE31-0504-01); Germany: Deutsche Forschungsgemeinschaft (DFG - 469666862); China: Research Grants Council (GRF); Italy: Ministero dell'Università e della Ricerca (NextGenEU 153D23001490006 M4C2.1.1, NextGenEU I53D23000820006 M4C2.1.1, NextGenEU I53D23001490006 M4C2.1.1, SOE2024\_0000023); Japan: Japan Society for the Promotion of Science (JSPS KAKENHI  JP25H0063, JSPS KAKENHI JP22H01227, JSPS KAKENHI JP22H04944, JSPS KAKENHI JP22KK0227, JSPS KAKENHI JP24K23939, JSPS KAKENHI JP24KK0251, JSPS KAKENHI JP25H00650, JSPS KAKENHI JP25H01291, JSPS KAKENHI JP25K01023); Norway: Research Council of Norway (RCN-314472); Poland: Polish National Science Centre (NCN 2021/42/E/ST2/00350, NCN OPUS 2023/51/B/ST2/02507, NCN OPUS nr 2022/47/B/ST2/03059, NCN UMO-2019/34/E/ST2/00393, UMO-2022/47/O/ST2/00148, UMO-2023/49/B/ST2/04085, UMO-2023/51/B/ST2/00920, UMO-2024/53/N/ST2/00869); Spain: Agència de Gestió d¿Ajuts Universitaris i de Recerca. (AGAUR - 2023 BP 00141), Ministry of Science and Innovation (RYC2019-028510-I, RYC2020-030254-I, RYC2021-031273-I, RYC2022-038164-I), Ministerio de Ciencia, Innovación y Universidades/Agencia Estatal de Investigaci\'on (PID2022-142604OB-C22); Sweden: Carl Trygger Foundation (Carl Trygger Foundation CTS 22:2312), Swedish Research Council (Swedish Research Council 2023-04654, VR 2021-03651, VR 2022-03845, VR 2022-04683, VR 2023-03403, VR 2024-05451, VR 2025-05940), Knut and Alice Wallenberg Foundation (KAW 2023.0366); Switzerland: Swiss National Science Foundation (SNSF - PCEFP2\_194658); United Kingdom: The Binks Trust, Royal Society (NIF-R1-231091); United States of America: U.S. Department of Energy (ECA DE-AC02-76SF00515), John Templeton Foundation (John Templeton Foundation 63206), Neubauer Family Foundation.

%
%


%
\clearpage
\printbibliography
\clearpage
 
\begin{flushleft}
\hypersetup{urlcolor=black}
{\Large The ATLAS Collaboration}

\bigskip

\AtlasOrcid[0000-0002-6665-4934]{G.~Aad}$^\textrm{\scriptsize 102}$,
\AtlasOrcid[0000-0001-7616-1554]{E.~Aakvaag}$^\textrm{\scriptsize 17}$,
\AtlasOrcid[0000-0002-5888-2734]{B.~Abbott}$^\textrm{\scriptsize 122}$,
\AtlasOrcid[0000-0002-0287-5869]{S.~Abdelhameed}$^\textrm{\scriptsize 118a}$,
\AtlasOrcid[0000-0002-1002-1652]{K.~Abeling}$^\textrm{\scriptsize 54}$,
\AtlasOrcid[0000-0001-5763-2760]{N.J.~Abicht}$^\textrm{\scriptsize 48}$,
\AtlasOrcid[0000-0002-8496-9294]{S.H.~Abidi}$^\textrm{\scriptsize 30}$,
\AtlasOrcid[0009-0003-6578-220X]{M.~Aboelela}$^\textrm{\scriptsize 44}$,
\AtlasOrcid[0000-0002-9987-2292]{A.~Aboulhorma}$^\textrm{\scriptsize 36e}$,
\AtlasOrcid[0000-0001-5329-6640]{H.~Abramowicz}$^\textrm{\scriptsize 155}$
% ...
\bigskip
\\

% ...
$^{17}$Department for Physics and Technology, University of Bergen, Bergen; Norway.\\
% ...
$^{30}$Physics Department, Brookhaven National Laboratory, Upton NY; United States of America.\\
% ...
$^{36}$$^{(a)}$Facult\'e des Sciences Ain Chock, Universit\'e Hassan II de Casablanca;$^{(b)}$Facult\'{e} des Sciences, Universit\'{e} Ibn-Tofail, K\'{e}nitra;$^{(c)}$Facult\'e des Sciences Semlalia, Universit\'e Cadi Ayyad, LPHEA-Marrakech;$^{(d)}$LPMR, Facult\'e des Sciences, Universit\'e Mohamed Premier, Oujda;$^{(e)}$Facult\'e des sciences, Universit\'e Mohammed V, Rabat;$^{(f)}$Institute of Applied Physics, Mohammed VI Polytechnic University, Ben Guerir; Morocco.\\
% ...
$^{44}$Physics Department, Southern Methodist University, Dallas TX; United States of America.\\
% ...
$^{48}$Fakult\"{a}t Physik, Technische Universit{\"a}t Dortmund, Dortmund; Germany.\\
% ...
$^{54}$II. Physikalisches Institut, Georg-August-Universit\"{a}t G\"ottingen, G\"ottingen; Germany.\\
% ...
$^{102}$CPPM, Aix-Marseille Universit\'e, CNRS/IN2P3, Marseille; France.\\
% ...
$^{118}$$^{(a)}$New York University Abu Dhabi, Abu Dhabi;$^{(b)}$United Arab Emirates University, Al Ain; United Arab Emirates.\\
% ...
$^{122}$Homer L. Dodge Department of Physics and Astronomy, University of Oklahoma, Norman OK; United States of America.\\
% ...
$^{155}$Raymond and Beverly Sackler School of Physics and Astronomy, Tel Aviv University, Tel Aviv; Israel.\\
% ...
$^{a}$ Also at Affiliated with an institute formerly covered by a cooperation agreement with CERN.\\
$^{b}$ Also at An-Najah National University, Nablus; Palestine.\\
$^{c}$ Also at Borough of Manhattan Community College, City University of New York, New York NY; United States of America.\\
$^{d}$ Also at Center for Interdisciplinary Research and Innovation (CIRI-AUTH), Thessaloniki; Greece.\\
$^{e}$ Also at Centre of Physics of the Universities of Minho and Porto (CF-UM-UP); Portugal.\\
$^{f}$ Also at CERN, Geneva; Switzerland.\\
% ...

\end{flushleft}


%

%
%
%
%

%
%
%
%

\end{document}